\documentclass[preprint,tightenlines,preprintnumbers,aps,eqsecnum,nofootinbib,superscriptaddress,showpacs,floatfix]{revtex4-1} 
\usepackage{amsmath,amssymb,amsbsy,amsfonts,bm}
\usepackage{graphicx}
\usepackage{bm}
\usepackage[mathscr]{euscript}
\usepackage{color}
\usepackage{rotating}
\usepackage{slashed}
\usepackage{float}
\usepackage{hyperref}

\usepackage{setspace}
\graphicspath{{./figures/}}

\newcommand\tr{\ensuremath{\mathop{\mathrm{tr}}}}

\newcommand{\sign}{\ensuremath{\mathop{\mathrm{sign}}}}

\newcommand{\order}{{\it O}}
\newcommand{\ba}{\begin{eqnarray}}
\newcommand{\ea}{\end{eqnarray}}
\newcommand{\be} {\begin{equation}}
\newcommand{\ee} {\end{equation}}

\newcommand{\cpt}{\raise0.4ex\hbox{$\chi$}PT}
\newcommand{\scpt}{S\raise0.4ex\hbox{$\chi$}PT}
\newcommand{\rscpt}{rS\raise0.4ex\hbox{$\chi$}PT}

\def\bea{\begin{eqnarray}}
\def\eea{\end{eqnarray}}

\begin{document}

\singlespacing

\preprint{FERMILAB-PUB-12-258-PPD}

\title{Neutral \boldmath$B$-meson mixing from three-flavor lattice QCD:                  
Determination of the $SU(3)$-breaking ratio $\xi$}

\author{A.~Bazavov}
\affiliation{Physics Department, Brookhaven National Laboratory, Upton, New York, USA}

\author{C.~Bernard}
\affiliation{Department of Physics, Washington University, St.~Louis, Missouri, USA}

\author{C.M.~Bouchard}
\affiliation{Physics Department, University of Illinois, Urbana, Illinois, USA}
\affiliation{Fermi National Accelerator Laboratory, Batavia, Illinois, USA}
\affiliation{Department of Physics, The Ohio State University, Columbus, Ohio, USA}

\author{C.~DeTar}
\affiliation{Physics Department, University of Utah, Salt Lake City, Utah, USA}

\author{M.~Di~Pierro}
\affiliation{School of Computing, DePaul University, Chicago, Illinois, USA}

\author{A.X.~El-Khadra}
\affiliation{Physics Department, University of Illinois, Urbana, Illinois, USA}

\author{R.T.~Evans}
\affiliation{Physics Department, University of Illinois, Urbana, Illinois, USA}
\affiliation{Department of Nuclear Engineering, North Carolina State University, Raleigh, North Carolina, USA}

\author{E.D.~Freeland}
\affiliation{Physics Department, University of Illinois, Urbana, Illinois, USA}
\affiliation{Department of Physics, Benedictine University, Lisle, Illinois, USA\
}

\author{E.~G\'amiz}
\email{megamiz@ugr.es}
\affiliation{Fermi National Accelerator Laboratory, Batavia, Illinois, USA}
\affiliation{CAFPE and Departamento de F\'{\i}sica Te\'orica y del Cosmos,
Universidad de Granada, Granada, Spain}

\author{Steven~Gottlieb}
\affiliation{Department of Physics, Indiana University, Bloomington, Indiana, USA}

\author{U.M.~Heller}
\affiliation{American Physical Society, Ridge, New York, USA}

\author{J.E.~Hetrick}
\affiliation{Physics Department, University of the Pacific, Stockton, California, USA}

\author{R.~Jain}
\affiliation{Physics Department, University of Illinois, Urbana, Illinois, USA}

\author{A.S.~Kronfeld}
\affiliation{Fermi National Accelerator Laboratory, Batavia, Illinois, USA}

\author{J.~Laiho}
\affiliation{SUPA, School of Physics and Astronomy, University of Glasgow, Glasgow,
Scotland, UK}

\author{L.~Levkova}
\affiliation{Physics Department, University of Utah, Salt Lake City, Utah, USA}

\author{P.B.~Mackenzie}
\affiliation{Fermi National Accelerator Laboratory, Batavia, Illinois, USA}

\author{E.T.~Neil}
\affiliation{Fermi National Accelerator Laboratory, Batavia, Illinois, USA}

\author{M.B.~Oktay}
\affiliation{Physics Department, University of Utah, Salt Lake City, Utah, USA}

\author{J.N.~Simone}
\affiliation{Fermi National Accelerator Laboratory, Batavia, Illinois, USA}

\author{R.~Sugar}
\affiliation{Department of Physics, University of California, Santa Barbara,
California, USA}
\author{D.~Toussaint}
\affiliation{Department of Physics, University of Arizona, Tucson, Arizona, USA}

\author{R.S.~Van~de~Water}
\affiliation{Physics Department, Brookhaven National Laboratory, Upton, New York, USA}

\collaboration{Fermilab Lattice and MILC Collaborations}
\noaffiliation

\date{\today}

\begin{abstract}

We study $SU(3)$-breaking effects in the neutral $B_d$-$\bar B_d$ and $B_s$-$\bar B_s$ 
systems with unquenched $N_f=2+1$ lattice QCD. 
We calculate the relevant matrix elements on the MILC collaboration's
gauge configurations with asqtad-improved staggered sea quarks.
For the valence light-quarks ($u$, $d$, and $s$) we use the asqtad action, 
while for $b$ quarks we use the Fermilab action.
We obtain $\xi = f_{B_s} \sqrt{B_{B_s}}                                                 
/ f_{B_d} \sqrt{B_{B_d}} = 1.268\pm 0.063$. We also present results
for the ratio of bag parameters $B_{B_s}/B_{B_d}$
and the ratio of CKM matrix elements $|V_{td}|/|V_{ts}|$.
Although we focus on the calculation of $\xi$, the strategy and techniques
described here will be employed in future extended studies of the $B$ mixing
parameters $\Delta M_{d,s}$ and $\Delta\Gamma_{d,s}$ in the Standard Model
and beyond.

\end{abstract}

\maketitle

\section{Introduction}

\label{introduccion}

The observation of new particles at high-energy colliders is not the only way 
for new physics to be discovered. It can also be unveiled  through the 
observation of deviations from the Standard Model (SM) via high-precision 
measurements of low-energy observables in high-luminosity experiments. 
This requires matching precision in the theoretical SM predictions for 
these observables. In principle, such a comparison could reveal the exchange 
of virtual, new heavy particles involving scales 
much higher than those that can be achieved in direct production at 
high-energy colliders.

\begin{figure}
\begin{minipage}[c]{.49\textwidth}
\includegraphics[width=0.95\textwidth]{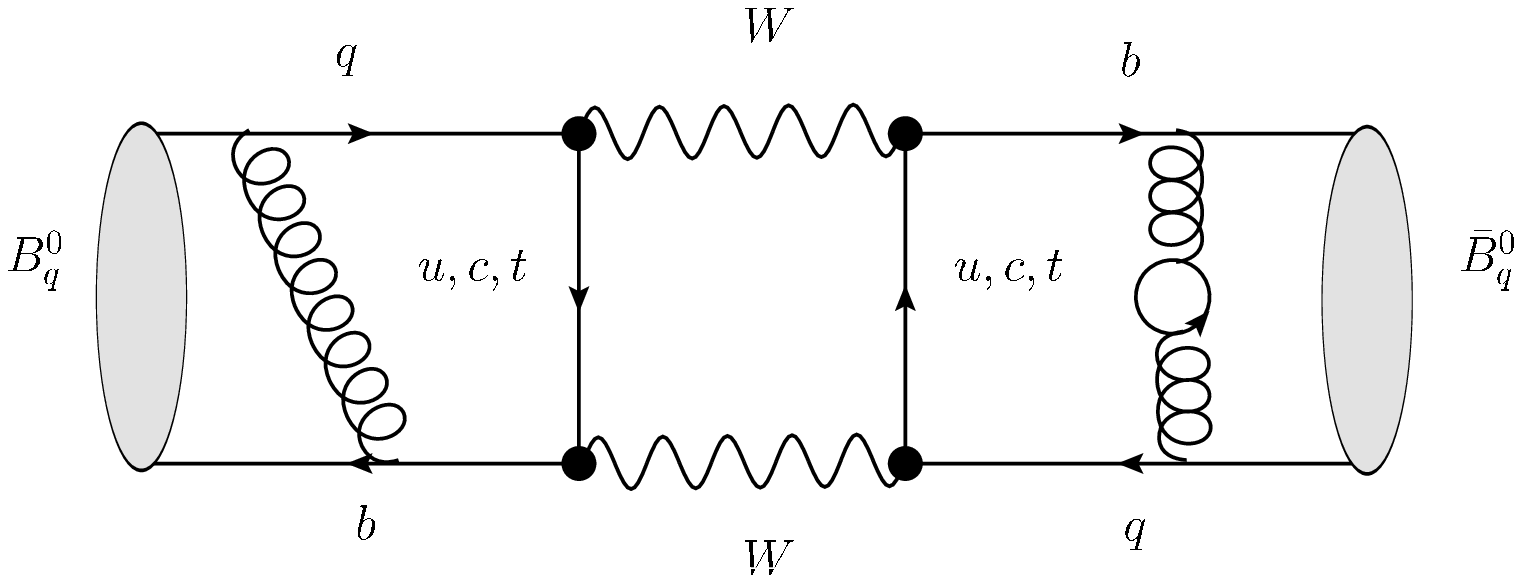}
\end{minipage}
\begin{minipage}[c]{.49\textwidth}

\includegraphics[width=0.95\textwidth]{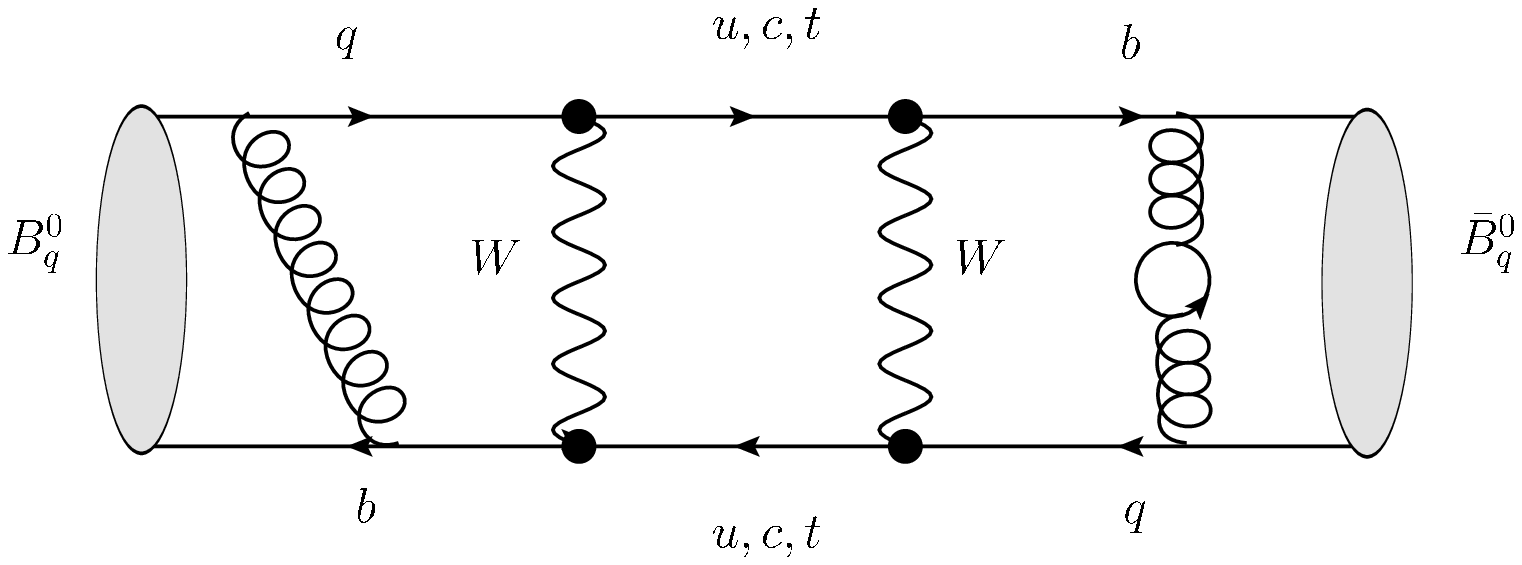}
\end{minipage}
\caption{Box diagrams contributing to $B^0-\bar B^0$ mixing in the SM. 
Gluon exchanges shown in the plot are just representative of the QCD corrections.  
\label{boxdiagams}}
\end{figure}

Heavy-flavor physics and, in particular, neutral-meson mixing are 
potentially very sensitive to these virtual effects. Neutral-meson 
mixing occurs at loop level in the SM, see 
Fig.~\ref{boxdiagams}, and it is further suppressed by small 
Cabibbo--Kobayashi-Maskawa (CKM) matrix elements, so the effect of new particles in the 
internal loops could be noticeable in the parameters describing the mixing. 
Indeed, there are several measurements for which there is a 2--3$\sigma$ 
difference from the SM prediction. These include 
$\sin(2\beta)$~\cite{sin2beta}, the like-sign dimuon charge 
asymmetry~\cite{D0dimuon}, and unitarity triangle (UT) 
fits~\cite{UTfits1,UTfits1bis,UTfits2,UTfits3,UTfits4}. It has been  
argued that these differences may be due to physics beyond the Standard Model (BSM)
affecting the neutral $B$-meson mixing processes~\cite{UTfits1,UTfits1bis}.

In the $B_s^0$ system, the relative phase between the decay amplitudes with 
and without mixing, $\beta_s$, could also show BSM effects, as pointed out in 
Ref.~\cite{lenznierste06} and later hinted at in a Tevatron 
measurement~~\cite{UTfit08}. Although new 
measurements at CDF~\cite{CDFbetas} and D\O~\cite{D0betas} are in better agreement 
with the SM, reducing the difference from $\sim 3\sigma$ to $\sim 1\sigma$, 
there is still room for a large deviation of $\beta_s$ from SM values. 

The main parameters describing mixing in the $B^0_s$ and the $B^0_d$ systems 
are the mass differences, $\Delta M_{s(d)}$, and the decay width 
differences, $\Delta \Gamma_{s(d)}$, between the heavy and light 
$B^0_{s(d)}$ mass eigenstates, and the $CP$ violating phases  
$\phi_{s(d)}$. The phases $\phi_{s(d)}$ are defined as the argument of the 
ratio of the dispersive and absorptive off-diagonal elements of the time 
evolution matrix which describes the 
mixing~\cite{mixingparam}. The existence of new, heavy particles in 
loops could affect the value of the mass differences, given by the dispersive 
part of the time evolution matrix. The mass differences
$\Delta M_{s}$~\cite{DeltaMsaverage1,DeltaMsaverage2,DeltaMsLHCb} and 
$\Delta M_{d}$~\cite{PDG2011} have been measured with an accuracy better than 1\%.
Improving the theoretical control on these quantities is thus crucial in order 
to fully exploit the potential of $CP$ violating observables to search for 
nonstandard physics. 
In addition, the theoretical calculation of BSM contributions to 
mixing and the experimental measurement of $B^0$ mixing parameters can 
help in constraining BSM parameters and understanding new physics~\cite{UTfits3}. 
Several recent studies have addressed that 
task~\cite{UTfits1,UTfits1bis,UTfits2,UTfits3,UTfits4,BSMstudies1,BSMstudies2,BSMstudies3,BSMstudies4,BSMstudies5,BSMstudies6,BSMstudies7,BSMstudies8}, 
finding that one of the main limitations to further constraining the parameter 
space in BSM theories is the error associated with 
the theoretical calculation of the nonperturbative inputs.

The most interesting quantity to analyze in $B^0$ mixing phenomena 
is the $SU(3)$-breaking ratio $\xi$, which measures the difference between
the mixing parameters in the $B^0_s$ and the $B^0_d$ systems, and enters the 
relation between the ratio of mass differences and CKM matrix elements as
\ba
\left\vert\frac{V_{td}}{V_{ts}}\right\vert =
\xi \sqrt{\frac{\Delta M_d M_{B_s}}
{\Delta M_s M_{B_d}}}\,.
\ea
Its value, together with the experimental measurement of the mass 
differences $\Delta M_{s,d}$, determines the ratio of CKM matrix elements 
$\vert V_{td}/V_{ts}\vert$, which constrains one side of the unitarity 
triangle~\cite{Wolfenstein83,BLO94}. Thus, $\xi$ is one of the key 
ingredients in UT analyses~\cite{UTfits1,UTfits2,UTfits3,UTfits4}. 

In the SM, mixing is due to box diagrams with the exchange of two $W$-bosons, 
like those in Fig.~\ref{boxdiagams}. 
These box diagrams can be rewritten in terms 
of an effective Hamiltonian with four-fermion operators describing 
processes with $\Delta B=2$. In BSM theories, mixing processes can receive 
contributions from additional diagrams due to the exchange of new,  
heavy particles. These can also be parametrized in terms of four-fermion 
effective operators built with SM degrees of freedom. 
The most general effective Hamiltonian describing processes 
with $\Delta B=2$ was given in~\cite{b2hamiltonian1,b2hamiltonian2}, 
and can also be found in \cite{B0proceedings2011}. There are a total of five 
independent operators (plus parity conjugates) in the Hamiltonian, but only 
three of them contribute to mixing in the SM 
\ba \label{effH}
{\cal H}_{\text{eff,SM}}^{\Delta B=2} = \sum_{i=1}^3 C_i {\cal O}_i\,, 
\ea
with
\ba\label{susybasis}
  {\cal O}_1^q & = &  \left(\bar q^i \gamma^\nu \,L\, b^i\right)
\left(\bar q^j\gamma^\nu \,L\, b^j\right)\,,\nonumber\\
   {\cal O}_2^q  & = &  \left (\bar q^i \,L\, b^i\right)\left(\bar q^j 
\,L\, b^j\right)\,,\nonumber\\
{\cal O}_3^q & = & \left (\bar q^i \,L\, b^j\right)\left(\bar q^j \,L\, 
b^i\right)\,,
\ea
where $i$ and $j$ are color indices, and $L$ and $R$ are the Dirac  
projection operators $\frac{1}{2}(1-\gamma_5)$ and $\frac{1}{2}
(1+\gamma_5)$ respectively. 
The fields $q$ denote strange or down fields for $B_s^0$ and $B_d^0$ 
mixing respectively, and $b$ represents the bottom field. 

The matrix element of the first operator in Eq.~(\ref{susybasis}), 
${\cal O}_1^q$, provides the mass difference in the SM: 
\ba\label{SMMsd}
\Delta M_{q}^{SM}=\frac{G_F^2M_W^2}{6\pi^2}
\vert V_{t q}^*V_{tb}\vert^2
\eta_2^BS_0(x_t)M_{B_{q}}
f_{B_{q}}^2\hat B_{B_{q}}\, ,
\ea
where $S_0(x_t)$ is the Inami-Lim function~\cite{InamiLim}, which depends on the top quark 
mass through $x_t=m_t^2/M_W^2$, and the quantity $\eta_2^B$ is a perturbative QCD 
correction factor. The products $f_{B_{q}}^2\hat B_{B_{q}}$ 
parametrize the hadronic matrix elements in the effective theory by
\ba
\label{relation}
\langle \bar B^0_q\vert {\cal O}_1^{q} \vert B^0_q
\rangle(\mu) = 
\frac{2}{3}M^2_{B_{q}}f^2_{B_{q}}B_{B_{q}}(\mu)
\, .
\ea
The factors $f_{B_{q}}$ are the $B_{q}^0$ decay constants.
The renormalization group invariant bag parameters 
$\hat B_{B_{q}}$ in Eq.~(\ref{SMMsd}) are related to the scheme 
and scale dependent bag parameters in (\ref{relation}) 
at next-to-leading order (NLO) by 
\ba
\hat B_{B_{q}} = \left[\alpha_s(\mu)\right]^{-6/23}
\,\left[1+\frac{\alpha_s(\mu)}{4\pi}J_5 \right]\,B_{B_{q}}(\mu)\, , 
\ea
where $J_5$ is known in both $\overline{{\rm MS}}$-NDR (naive dimensional
regularization) and  $\overline{{\rm MS}}$-HV ('t~Hooft-Veltman) schemes 
\cite{BJW90}. 
Bag parameters have traditionally been used to measure the deviation 
of the four-fermion operator matrix elements from their vacuum insertion 
values, $B_B=1$.

The $SU(3)$-breaking parameter $\xi$ can be written in terms of decay 
constants and bag parameters 
\ba
\xi = \frac{f_{B_s}\sqrt{B_{B_s}}}{f_{B_d}\sqrt{B_{B_d}}}\,.
\ea
Many of the uncertainties that affect the theoretical calculation
of the decay constants and bag parameters cancel totally or partially
in this ratio, leaving the chiral extrapolation as the dominant error. 
Hence, $\xi$ and the combination of CKM matrix elements
related to it, can be determined with a significantly
smaller error than the individual matrix elements.

The hadronic matrix elements in Eq.~(\ref{relation}) encode the 
nonperturbative physics of the problem and are best calculated using 
lattice QCD. Our current knowledge of them limits the accuracy with 
which the CKM matrix elements 
appearing in Eq.~(\ref{SMMsd}) can be determined from the experimental 
measurements of $\Delta M_{s(d)}$. In particular, the
uncertainty associated with the calculation of $\xi$ is one of the main
limiting factors in UT analyses, so improvement in the knowledge of $\xi$ 
is crucial to disentangle the origin of the $2$--$3\sigma$ tension.

There are two $2+1$ unquenched lattice calculations of the ratio $\xi$ in the 
literature. One is by the HPQCD collaboration~\cite{HPQCB0mixing}, which quotes 
the value $\xi=1.258(33)$. The other is an exploratory study by the RBC and 
UKQCD collaborations~\cite{RBCxi}  
on a single lattice spacing and using the static limit for the bottom quark; 
their result is $\xi=1.13(12)$. In this paper, we report a lattice 
calculation of $\xi$ at the few percent level. 

Preliminary results related to the work here were presented in
\cite{lattice06,lattice07,lattice08,lattice09}. In Ref.~\cite{lattice06}, the
simulation and correlator fitting methods were described using data for
one lattice spacing, while Refs.~\cite{lattice07,lattice08} focused on the 
discussion of statistical and fitting errors, and the chiral extrapolation 
method. In Ref.~\cite{lattice09} we studied the matching method and the 
heavy-quark discretization errors.

The primary difference between this work and the HPQCD calculation in 
Ref.~\cite{HPQCB0mixing} is the treatment of the valence $b$ quarks. The HPQCD 
collaboration uses lattice NRQCD~\cite{nrqcd} while we employ the clover 
action~\cite{clover} with the Fermilab interpretation~\cite{Fermilab}. 
An advantage of the Fermilab method is that it can also be 
efficiently used to simulate charm quarks, so the analysis performed
in this work can be easily extended to the study of the short-distance
contributions to $D^0$-$\bar D^0$ mixing. Although in the case of neutral
$D$ mixing the long-distance contributions are believed to be dominant,   
a calculation of the short-distance contributions nevertheless can provide 
valuable constraints on extensions of the SM~\cite{GHPP07}.

In order to achieve the few-percent level of precision required by 
phenomenology, we use lattice QCD simulations with realistic sea quarks. 
In particular, we employ a subset of the MILC configurations  with 2+1 
flavors of asqtad sea quarks 
\cite{bernard-2001-64,alford-1995-361,1985PhLB..158..250L}. In the valence 
sector, we use the same staggered asqtad action to simulate the light quarks. 
The configurations we use in this analysis were generated using the 
fourth-root procedure for eliminating extra degrees of freedom originating  
from fermion doubling. Despite the nonlocal violations of unitarity of the 
rooted theory at non-zero lattice spacing~\cite{Prelovsek05,BGS06}, there 
are strong theoretical arguments~\cite{Bernard06,Shamir04,Shamir06,BGS08}, 
as well as other analytical and numerical 
evidence~\cite{SharpePoS06,KronfeldPoS07,GoltermanPoS08,Donaldetal11},  
that the local, unitary theory of QCD is recovered in the continuum 
limit. This gives us confidence that the rooting procedure yields valid results. 
We also explicitly tested the rooting procedure as well as improvements 
in our heavy action by calculating the spin-dependent hyperfine 
splittings for $B_s$ and $D_s$ mesons in Ref.~\cite{tuning10}.

Our collaboration has already successfully used the asqtad MILC ensembles in similar 
calculations of other quantities involving $B$ mesons, as part of a 
broad program of calculating matrix elements: 
for example, the extraction of the CKM matrix elements $\vert V_{ub}\vert$ and
$\vert V_{cb}\vert$ from the calculation of, respectively, the semileptonic form
factors describing the processes $B\to\pi l \nu$ \cite{RuthBtopi} and
$B\to D^* l \nu$ \cite{JackBtoD*1,JackBtoD*2}; or, more recently, the
calculation of the $f_B$ and $f_{B_s}$ decay 
constants~\cite{decayconstants2011} and the form-factor ratios between 
the semileptonic decays $\bar B\to D^+l^-\bar\nu$ and  
$\bar B_s\to D_s^+l^-\bar\nu$~\cite{Bsemileptonicratio}.

This paper is organized as follows. In Sec.~\ref{sec:numerical}, we describe 
the actions and parameters used in our numerical simulations, as well as the 
construction of the mixing operators and correlation functions. 
Section~\ref{sec:matching} presents the renormalization method using one-loop 
mean-field improved lattice perturbation theory. We include 
a discussion of the errors associated with 
the matching and numerical values of the matching coefficients used. Next, 
in Sec.~\ref{sec:fitting}, we give the details of the procedure for the 
correlator fits. Section~\ref{sec:chiralfits} is devoted to the 
chiral-continuum extrapolation, which is performed within the framework of 
rooted staggered chiral perturbation theory~~\cite{SharpeandLee,aubin-2003,Aubin:2003uc,sharperuth04,aubin-2006-73}. We describe and discuss 
the choice of the functional form used in the extrapolation, the different 
fitting methods tested, and the choice 
of parameters and parametrization. In Sec.~\ref{sec:Errors}, we list and 
estimate the different systematic errors. Finally, 
Section~\ref{sec:Results} compiles our final results for the parameter $\xi$ 
as well as for $\vert V_{td}\vert/\vert V_{ts}|$, and the ratio of bag 
parameters $B_{B_s}/B_{B_d}$. We also discuss planned future improvements 
in the calculation of $B^0$ mixing parameters by our collaboration. 
In Appendix~\ref{app:SCHPT}, we provide the explicit formulas for the chiral 
fit functions used in the chiral fits described in 
Section~\ref{sec:chiralfits}. In Appendix~\ref{app:HQDerrors}, we compile 
the functions needed to estimate the heavy-quark discretization errors in 
our calculation. Finally, Appendix~\ref{app:priors} discusses our choices 
for prior central values and widths for the correlator fits.  

\section{Numerical Simulations}
\label{sec:numerical}

\begin{table}[t]
    \centering
\caption{\label{tab:ensemblesa} 
Parameters of the ensembles analyzed in this work.
The first two rows show the approximate lattice spacing and the volume.
$am_l$ and $am_h$ are the light and strange sea quark masses, respectively.
$N_{{\rm confs}}$ is the number of configurations analyzed from each ensemble, and 
$am_q$ are the light valence quark masses. The $r_1/a$
values are obtained by fitting the calculated $r_1/a$ to a smooth
function~\cite{Allton96}, as explained in Ref.~\cite{MILCasqtad}.}
  \begin{tabular}{cccccc}
\hline
\hline 
$\approx a({\rm fm})$ & $\left(\frac{L}{a}\right)^3\times\frac{T}{a}$ & 
$am_l/am_h$ & $N_{{\rm confs}}$ & $am_q$ & $r_1/a$\\
\hline
0.12 & $24^3\times 64$ & 0.005/0.05 & 529 & 0.005, 0.007, 0.01, 0.02, 0.03, 
0.0415 & 2.64 \\
0.12 & $20^3\times 64$ & 0.007/0.05 & 833 & 0.005, 0.007, 0.01, 0.02, 0.03, 
0.0415 & 2.63 \\
0.12 & $20^3 \times 64$ & 0.01/0.05 & 592 & 0.005, 0.007, 0.01, 0.02, 0.03, 
0.0415 &  2.62 \\
0.12 & $20^3 \times 64$ & 0.02/0.05 & 460 & 0.005, 0.007, 0.01, 0.02, 0.03, 
0.0415 & 2.65 \\
\hline
0.09 & $28^3 \times 96$ & 0.0062/0.031 & 557 & 0.0031, 0.0044, 0.062, 0.0124, 
0.0272, 0.031 & 3.70 \\
0.09 & $28^3 \times 96$ & 0.0124/0.031 &  534 & 0.0031, 0.0042, 0.062, 0.0124, 
0.0272, 0.031 & 3.72 \\
\hline\hline
\end{tabular}
\end{table}

\subsection{Parameters of the simulations}

The $n_f = 2+1$ MILC ensembles~\cite{MILCasqtad} used in our calculation include 
the effect of three sea-quark flavors: two degenerate light quarks corresponding 
to the up and down quarks (although with larger masses than the physical ones), and 
one heavier quark corresponding to the strange quark. These dynamical quarks are 
simulated using the asqtad improved staggered action with errors  starting 
at $O(\alpha_s a^2)$~\cite{lepage-1999-59}. The gluon action is a
Symanzik improved and tadpole improved action, with $O(\alpha_s a^2)$  errors 
coming from the gluon loops removed~\cite{1983NuPhB.226..205S,lepage-1993-48}. 
The couplings needed to remove the $O(\alpha_s a^2)$ errors coming from quark  
loops~\cite{alphasa207} were available only after the generation of configurations 
was well advanced, so these effects are not accounted for in the MILC ensembles. 
The dominant errors in the gauge action are thus also of $O(a^4, \alpha_s a^2)$.

The valence light-quark propagators are generated using the asqtad action 
and converted 
to naive quark propagators using the relation~\cite{wingate-2003-67}
\be\label{eq:naive}
S_{{\rm naive}}(x,y)=\Omega(x) \Omega^{\dagger}(y)S_{{\rm staggered}}(x,y).
\ee
where $\Omega(x)=\gamma_0^{x_0}\gamma_1^{x_1}\gamma_2^{x_2}\gamma_3^{x_3}$.

For the  heavy bottom quarks we 
use the Sheikholeslami-Wohlert action \cite{clover} 
with the Fermilab interpretation for heavy-quark systems
\cite{Fermilab}.  This interpretation retains the full mass dependence 
of the theory within the parameters of the lattice action.  A tree-level 
matching to QCD is then performed via heavy quark effective theory (HQET), 
after which it can be shown that the errors in the action begin at 
$\order(\alpha_s\Lambda_{{\rm QCD}}a,\Lambda_{{\rm QCD}}^2a^2)$ times bounded 
functions of $m_ba$, the $b$-quark mass in lattice units. 

We perform our analysis at two different values of the lattice spacing, 
$a\approx 0.12,0.09~{\rm fm}$, and for a variety of sea-quark masses. The 
values used are shown in Table~\ref{tab:ensemblesa}. The mass of the heavy 
$b$ quark is fixed to its physical value by computing the spin-averaged 
$B_s$ kinetic mass \cite{tuning10}. This determines the $b$ quark's 
hopping parameters,  $\kappa_b=0.0860$ for the $a \approx 0.12~{\rm fm}$  
lattice and  $\kappa_b=0.0923$ for the $a \approx 0.09~{\rm fm}$ 
lattice \cite{tuning10}, and thus the bare $b$ quark mass. 
We simulate the $B$ mesons with the six different values of 
light-valence quark mass listed in Table~\ref{tab:ensemblesa}, the smallest 
of which is around $m_s/8$, in order to facilitate the 
extrapolation/interpolation to the physical down/strange quark masses.

\subsection{Correlators: the open-meson propagator} \label{sec:open}

\begin{figure}[tb]
\vspace*{-2.cm}                                                                              
\includegraphics[scale=0.8]{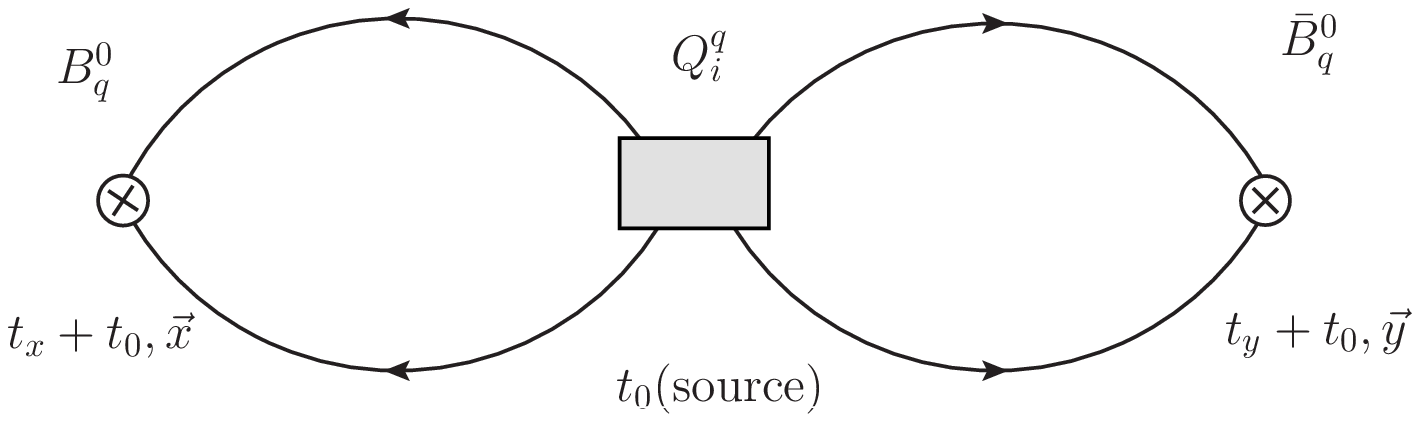}
\vspace*{-18.cm}
\caption{\label{fig:corr_diagram} Structure of the three-point correlators. A $B^0_q$ is
created at rest at $t_x+t_0<t_0$. At time $t_0$, it oscillates into a $\bar B^0_q$ via
the operator ${\cal O}_i^q$, which is subsequently annihilated at $t_y+t_0>t_0$.}
\end{figure}

As described in the introduction, the study of the $SU(3)$-breaking ratio 
$\xi$ requires the calculation of the hadronic matrix elements\footnote{To simplify 
the notation, we define $\langle{\cal O}_i^q\rangle \equiv 
\langle\bar B^0_{q}\vert {\cal O}_i^q \vert  B^0_{q} \rangle$ 
for $i=1,2$.} 
$\langle{\cal O}_1^q\rangle$ and $\langle{\cal O}_2^q\rangle$, the latter of  which 
mixes with $\langle{\cal O}_1^q\rangle$ under renormalization, 
for both $q=d,s$. The matrix elements 
are obtained from three-point correlation functions with zero spatial momentum
\be \label{eqn:corrQ}
C_{\mathcal{O}_i^q}(t_x,t_y) = \sum_{\bm{x}, \bm{y}} \langle \bar{B}^0_q(t_y,\bm{y}) 
\mathcal{O}_i^q(0) B^0_q(t_x,\bm{x})^\dagger\rangle\,,
\ee
where $i=1$ or $2$, and the $B$-meson creation operator $B_q^0(t,\bm{x})^\dagger=
\sum_{\bm{x'}}\bar b(t,\bm{x'})S(\bm{x},\bm{x'})\gamma_5 q(t,\bm{x})$, 
with $q(t,{\bm x})$ the 
naive light quark field, whose propagator is constructed from the staggered 
propagator {\it via} Eq.~(\ref{eq:naive}), and with $S(x,x')$ a smearing 
function. Our choice of smearing function is discussed in Sec.~\ref{fitting}.  
The structure of the 
functions in (\ref{eqn:corrQ}) is depicted in Fig.~\ref{fig:corr_diagram}. 
The four-fermion operators 
${\cal O}_i^q$ are placed at the origin while $B$-mesons are positioned at 
$x$ and $y$. This layout allows us to perform the three-point function fits over 
both $t_x$ and $t_y$, maximizing the information included in the fits. 
In order to extract the relevant matrix elements from (\ref{eqn:corrQ}), we need to
determine the overlap of the $B$-meson creation operator with the ground state.  
Therefore, we also  need the pseudoscalar two-point correlator with zero 
spatial momentum
\be \label{eqn:corr2pt}
C_{PS} (t) = \sum_{\bm{x}} \langle B^0_q(t,\bm{x})
B^0_q(0,\bm{0})^\dagger\rangle\,.
\ee

The calculation of both three-point and two-point correlators can be 
organized into convenient structures.  
Starting with a general correlator with Dirac structure 
$\Gamma_1\times\Gamma_2$, which accommodates a full set of $\Delta B=2$ 
four-quark operators, including those in Eq.~(\ref{susybasis}),
\be
C_3(t_x,t_y) = \sum_{\bm{x},\bm{y}} \langle \bar{B}^0_q(t_y,\bm{y}) 
\bar{q}(0)\Gamma_1 b(0) \bar{q}(0)\Gamma_2 b(0) B^0_q(t_x,\bm{x})^\dagger 
\rangle\,,
\ee
and performing the four possible Wick contractions, we obtain
\begin{eqnarray}\label{eq:wickcont}
C_3(t_x,t_y)=\sum_{\bm{x},\bm{y}}&&
\Big\{ \tr[\gamma_5 L_q(x,0)\Gamma_1 H_b(0,x)]\tr[\gamma_5 L_q(y,0)
\Gamma_2 H_b(0,y)] \nonumber\\&&+ \tr[\gamma_5 L_q(y,0)
\Gamma_1 H_b(0,y)]\tr[\gamma_5 L_q(x,0)\Gamma_2 H_b(0,x)] \nonumber \\ 
\nonumber &&-\tr[\gamma_5 L_q(x,0)\Gamma_1 H_b(0,y)\gamma_5 L_q(y,0)
\Gamma_2 H_b(0,x)]\nonumber\\&&-\tr[\gamma_5 L_q(x,0)\Gamma_2 
H_b(0,y)\gamma_5 L_q(y,0)
\Gamma_1 H_b(0,x)] \Big\} \\\nonumber = \sum_{\bm{x},\bm{y}}&&
\Big\{ \tr[L_q(x,0)\Gamma_1 \gamma_5 H_b^{\dagger}(x,0)]\tr[L_q(y,0)
\Gamma_2 \gamma_5 H_b^{\dagger}(y,0)]\nonumber\\
&&+\tr[L_q(y,0)\Gamma_1\gamma_5 
H_b^{\dagger}(y,0)]\tr[L_q(x,0)\Gamma_2 \gamma_5 H_b^{\dagger}(x,0)]\\
 \nonumber &&-\tr[L_q(x,0)\Gamma_1 \gamma_5 H_b^{\dagger}(y,0) 
L_q(y,0)\Gamma_2\gamma_5 H_b^{\dagger}(x,0)]\nonumber\\&&-\tr[ L_q(x,0)\Gamma_2 
\gamma_5 H_b^{\dagger}(y,0) L_q(y,0)\Gamma_1 \gamma_5 H_b^{\dagger}(x,0)]
\Big\},
\end{eqnarray}
where $L_q$ is the (naive) light-quark propagator, and $H_b$ is the heavy-quark 
propagator. The traces in Eq.~(\ref{eq:wickcont}) run over spin and color indices. 

These correlators can be rewritten as 
\ba \label{eq:3fromopen}
C_3(t_x,t_y)&=&\Gamma_1^{\beta\alpha}E^{\alpha\beta}_{aa}(t_x)\Gamma_2^{\tau\sigma}
E^{\sigma\tau}_{cc}(t_y)+
\Gamma_1^{\beta\alpha}E^{\alpha\beta}_{aa}(t_y)\Gamma_2^{\tau\sigma}E^{\sigma\tau}_{cc}
(t_x)\\ \nonumber&&-
\Gamma_1^{\beta\alpha}E^{\alpha\sigma}_{ac}(t_x)\Gamma_2^{\sigma\tau}E^{\tau\beta}_{ca}(t_y)
-\Gamma_1^{\beta\alpha}E^{\alpha\sigma}_{ac}(t_y)\Gamma_2^{\sigma\tau}
E^{\tau\beta}_{ca}(t_x)\,,
\ea
where summation over repeated indices is implied, and we have introduced 
the basic objects
\ba\label{eq:openmeson}
E^{\alpha\beta}_{ac}(t)=\gamma_5^{\alpha\sigma}H_{b,da}^{*\tau\sigma}(t,0)           
L_{q,dc}^{\tau\beta}(t,0)\,,
\ea
with Dirac indices labeled as $\alpha,\beta,\sigma,\tau$ and color
indices labeled as $a,c,d$. We call the combination of propagators 
$E^{\alpha\beta}_{ad}(t_x)$ defined in Eq.~(\ref{eq:openmeson}) 
``open-meson propagator''. Once the open-meson propagators have been computed 
and saved, all correlation functions needed for $B$-meson mixing, including 
BSM operators, can be immediately constructed by contracting them with the 
appropriate Dirac structures. 
As shown in Eq.~(\ref{eq:3fromopen}) the three-point correlators are obtained 
by combining two open-meson propagators, while for the two-point correlators 
we only need one open-meson propagator.

\subsection{Doubler modes' effect on the correlation functions}

\label{sec:doubling}

The remnant doubling degeneracy of staggered fermions leads to
contributions of  scalar states, in addition to pseudoscalar states, 
in correlation functions with external pseudoscalar particles. 
The scalar contamination yields oscillating terms in the correlation 
functions~\cite{wingate-2003-67}. In this section, we extend the 
analysis of  Ref.~\cite{wingate-2003-67}, for two-point correlation functions, 
to the three-point functions introduced in Section~\ref{sec:open}.  
We conclude that the effect of the doubler modes on the three-point 
functions can be removed at leading order in the lattice spacing 
through appropriate fits of the Euclidean time-dependence.

The \emph{doubling} symmetry of the original naive action under the
transformation
\begin{equation}
\psi(x)\to e^{ix \cdot \pi_g} M_g \psi(x)\,, \qquad \bar{\psi(x)}
\to e^{ix\cdot \pi_g} \bar\psi(x) M_g^{\dagger}\,,
\end{equation}
where 
\begin{eqnarray}
M_g & = & \prod_{\mu \in g}i\gamma_5\gamma_{\mu}\,,\\
G& =& \{ g:g=(\mu_1,\mu_2,\ldots),\mu_1<\mu_2<\ldots \},\\
(\pi_g)_{\mu}& =& \left \{ \begin{array} {cc}
\frac {\pi} a & \mbox{ if $\mu \in g$} \\
0 & \mbox{ otherwise}
    \end{array} \right.
\end{eqnarray}
generates sixteen equivalent species of quarks, referred to as \emph{tastes}, 
that can be reduced to four by staggering the quark field \cite{staggered}. 
Each  element of $G$ is a list of up to four indices, e.g., (2), (0,3), and (0,1,2,3) 
are elements of $G$, as is the empty set $\emptyset$. 
Different $g$'s label different doubler modes, or tastes. 

Consider the general three-point function in momentum space,
\ba\label{eq:firstC}
&&C_{\Gamma_1\times\Gamma_2}(t_x,t_y)\equiv\sum_{\bm{x},\bm{y}}\langle\bar{b}(x)
\gamma_5q(x)\left[ \bar{q}(0)
\Gamma_1 b(0) \bar{q}(0) \Gamma_2 b(0) \right] \bar{b}(y)\gamma_5 q(y)\rangle=
\nonumber\\
&&\int_{-\pi/a}^{\pi/a}\frac{ d^3\bm{p}} {(2\pi)^3} \frac{ d^3\bm{k}}
{(2\pi)^3}\langle \bar{\tilde{b}}(\bm{p},t_x)\gamma_5 \tilde{q}
(\bm{p},t_x) \left[ \bar{q}(0) \Gamma_1 b(0) \bar{q}(0) \Gamma_2 b(0) \right]
\bar{\tilde{b}}(\bm{k},t_y)\gamma_5 \tilde{q}(\bm{k},t_y)\rangle\,,
\ea
where $\Gamma_1\times \Gamma_2$ denotes the Dirac structure of the four-fermion
operators in (\ref{susybasis}). For simplicity of notation, we omit the smearing function 
from the $B$ meson operator and write it as $\bar{b}(x) \gamma_5 q(x)$. It would be 
straightforward (but not particularly instructive) to generalize the expressions of 
Eqs.~(\ref{eq:firstC})--(\ref{oscillations}) to include the smearing function.

For now the bracketed four-quark operator is left in position space and
$\tilde{b},\tilde{q}$ are the spatial momentum-space bottom and strange/down
fermion fields.  Because of the doubling symmetry, we can integrate over the
central half of the Brillouin zone and sum over the spatial doublers
\ba
C_{\Gamma_1\times\Gamma_2}(t_x,t_y) = 
\sum_{g_s, g'_s}\int_{-\pi/2a}^{\pi/2a}\frac{ d^3\bm{p}} {(2\pi)^3}
\frac{ d^3\bm{k}} {(2\pi)^3}&&\langle \bar{\tilde{b}}(\bm{p}+\pi_{g_s},t_x)
\gamma_5 \tilde{q}(\bm{p}+\pi_{g_s},t_x) \left[ \bar{q}(0) \Gamma_1 b(0)
\bar{q}(0) \Gamma_2 b(0) \right] \times \nonumber\\ &&\bar{\tilde{b}}(\bm{k}
+\pi_{g'_s},t_y)\gamma_5 \tilde{q}(\bm{k}+\pi_{g'_s},t_y)\rangle,
\ea
where $g_s$ denotes a particular spatial doubler mode.  Due to the high
momentum that is imparted to the heavy quark when $g_s\ne \emptyset$,
such states are far off-shell and have a negligible effect on the
correlation function. The taste of the temporal modes can now be considered
by Fourier transforming the light quarks' temporal component, and then again
restricting the Brillouin zone and summing over the doublers
\ba
C_{\Gamma_1\times\Gamma_2}(t_x,t_y)& = & 
\int_{-\pi/2a}^{\pi/2a}\frac{ d^3\bm{p}} {(2\pi)^3} \frac{ d^3\bm{k}}
{(2\pi)^3} \int_{-\pi/2a}^{\pi/2a}\frac{ d p_0} {(2\pi)}\frac{ d k_0}
{(2\pi)}e^{ip_0 t_x+ik_0 t_y} \\ \nonumber &&\times \Big\langle \bar{\tilde{b}}
(\bm{p},t_x)\gamma_5\left[ \tilde{q'}(\bm{p},p_0)+(-1)^{t_x}\tilde{q'}
(\bm{p},p_0+\pi/a)\right] \\ \nonumber &&\times \left[ \bar{q}(0) \Gamma_1 b(0)
\bar{q}(0) \Gamma_2 b(0) \right] \bar{\tilde{b}}(\bm{k},t_y)\gamma_5 \left[
\tilde{q'}(\bm{k},k_0)+(-1)^{t_y}\tilde{q'}(\bm{k},k_0+\pi/a)\right]
\Big\rangle.
\ea
With the momentum space spinors $\tilde{f'}^g$ defined as
\be
\tilde{f'}^g(p)=\sum_{\mu\in g} i\gamma_5 \gamma_{\mu}\tilde{q'}(p+\pi_g)
\ee
so that
\ba
\tilde{q'}(\bm{p},p_0)=\tilde{f'}(\bm{p},p_0),\quad \tilde{q'}(\bm{p},p_0+
\pi/a)=i\gamma_5 \gamma_0 \tilde{f'}^0(\bm{p},p_0),
\ea
the three-point function can be written as
\ba
C_{\Gamma_1\times\Gamma_2}(t_x,t_y) = 
\int_{-\pi/2a}^{\pi/2a}\frac{ d^3\bm{p}} {(2\pi)^3} \frac{ d^3\bm{k}}
{(2\pi)^3}\Big\langle \bar{\tilde{b}}(\bm{p},t_x)\gamma_5 \left[ \tilde{f}
(\bm{p},t_x)+(-1)^{(t_x)}i\gamma_5\gamma_0 \tilde{f}^{0}(\bm{p},t_x)\right]
\nonumber \\ \times \left[ \bar{q}(0) \Gamma_1 b(0) \bar{q}(0) \Gamma_2 b(0)
\right] \bar{\tilde{b}}(\bm{k},t_y)\gamma_5 \left[ \tilde{f}(\bm{k},t_y)
+(-1)^{(t_y)}i\gamma_5\gamma_0 \tilde{f}^0(\bm{k},t_y)\right]\Big\rangle\,,
\ea
where the superscript 0 indicates a temporal taste and no superscript is the
null taste at the center of the Brillouin zone.

After Fourier transforming, the bracketed four-quark operator has no restrictions
on the tastes that contribute to it.  However, it must be contracted with the
external quark fields to form the propagators.  Because the asqtad action is
used, contractions between tastes of different types are suppressed to
$\order(a^2\alpha_s^2)$.  
The three-point function then takes the form
\ba\label{oscillations}
C_{\Gamma_1\times\Gamma_2}(t_x,t_y)   & = &  
\int_{-\pi/2a}^{\pi/2a}\frac{ d^3\bm{p}} {(2\pi)^3} \frac{ d^3\bm{k}}
{(2\pi)^3}\Big\langle \bar{b}(\bm{p},t_x)\gamma_5 \left[ f(\bm{p},t_x)
+(-1)^{t_x}i\gamma_5\gamma_0 f^{0}(\bm{p},t_x)\right]\nonumber \\ &&
\times \left[ \left(\bar{f}(0)+i\gamma_5\gamma_0 \bar{f}^0(0)\right) \Gamma_1
b(0)\left(\bar{f}(0)+i\gamma_5\gamma_0\bar{f}^0(0)\right) \Gamma_2 b(0) \right]
\bar{b}(\bm{k},t_y)\gamma_5 \nonumber\\ &&\times \left[ f(\bm{k},t_y)
+(-1)^{t_y}i\gamma_5\gamma_0 f^0(\bm{k},t_y)\right]\Big\rangle\,,
\ea
where higher order terms coming from contractions between quarks of 
different taste give terms of $O(a^2 \alpha^2_s)$ that are not considered 
here.  The effects of such terms are comparable to NLO terms in staggered 
chiral perturbation theory and need to be considered at that order.  
They give rise to the ``wrong-spin'' terms discussed below. According to 
Eq.~(\ref{oscillations}), the leading-order correlation functions have 
contributions from both the pseudoscalar and the scalar
states. The latter ones are known as oscillating states, since the sign
of their contribution oscillates with time.

The fit ansatz for our correlators must model both regular and 
oscillating contributions,  so that we can remove the latter and extract 
the physical matrix elements. This is done using the  form
\be \label{eqn:CQ}
C_{{\cal O}_i^q}(t_x,t_y)=\sum_{\alpha,\beta=0}^{N{{\rm states}}-1} 
Z_\alpha Z_\beta \frac{O_{\alpha\beta}^i}{\sqrt{2E_\alpha 2E_\beta}}
\,(-1)^{(t_x+1)\alpha+(t_y+1)\beta\,}
e^{-E_{\alpha}t_x-E_\beta t_y}\,,
\ee
where the sum is over a finite number of states $N_{\text{states}}$. 
The time $t_x$ in Eq.~(\ref{eqn:CQ}) and in the discussion on fitting in 
Sec.~\ref{sec:fitting} is the number of time slices between the initial state and 
the operator, and thus it is a positive number, unlike the time $t_x$ 
defined in Fig.~\ref{fig:corr_diagram}. 
The oscillations in Euclidean time given by the factor 
$(-1)^{(t_x+1)\alpha+(t_y+1)\beta}$ reflect the contribution from the scalar 
states in Eq.~(\ref{oscillations}). The matrix elements of interest 
are given by the three-point amplitude of the ground state 
$\alpha=\beta=0$, $O_{00}^i$. Analogously, we incorporate regular and oscillating 
contributions to the description of the two-point correlators by using 
the following functional form in the fits
\be \label{eqn:CPS}
C_{PS}(t)=\sum_{\alpha=0}^{N_{{\rm states}}-1} |Z_\alpha|^2
\,(-1)^{(t+1)\alpha\,} \left(e^{-E_\alpha t}+e^{-E_\alpha (T-t)}\right)\,,
\ee
where $T$ is the temporal size of the lattice. 
Three- and two-point functions are fit simultaneously in our analysis, as 
described in Sec.~\ref{sec:fitting}. 

\subsection{Improving the heavy-light four-quark operator}

\label{sec:openmeson}

In addition to the discretization errors in the heavy-quark 
action, the mixing operator also has discretization errors 
due to the difference in the small-momentum behavior of lattice and continuum 
heavy quarks. In this section, we describe how the lowest order of operator 
discretization errors are removed in our calculation. We first show that the 
errors start at $\order(a\bm{p})$ and then discuss how the error at this order 
can be removed by a ``rotation'' of the heavy-quark field.
 
To begin, consider the small $a\bm{p}$ expansion of the spinor for the 
Wilson-like fermion
\begin{eqnarray} 
u^{{\rm lat}}(\chi,\bm{p})=\frac{\gamma_0 \sign\chi\sinh Ea-i\gamma_j
\sin(p_j a )+ L} 
{\sqrt{2L(L+\sinh Ea)}} u(\chi,0) \nonumber\\
=e^{-m_1a/2}\left[1-\frac {i\bm{\gamma}\cdot\bm{p}a}{2\sinh m_1a}+O
((a\bm{p})^2)\right]u(\chi,0),
\end{eqnarray}
where $\chi$ labels spin and particle {\it vs.} antiparticle, 
$\hat{p}=(2/a)\sin(pa/2)$, and for the clover action $L=1+m_0a+\frac 12 
\hat{\bm{p}}^{\,2}a^2-\cos p_0a$. The continuum spinor has the expansion
\be
u^{{\rm cont}}(\chi,\bm{p})=\left[1-\frac {i \bm{\gamma}\cdot\bm{p}} {2m}
+O((a\bm{p})^2)\right]u(\chi,0).
\ee
The mismatch between the small-momentum terms can be easily removed by 
``rotating'' the lattice heavy quark as was done to heavy-light bilinear 
operators in Ref.~\cite{Fermilab}. The light lattice spinors for the staggered 
formulation have the same small-momentum behavior as in the continuum up to 
$\order((\bm{p}a)^2,(m_qa)^2)$ and need not be matched. 

The analysis of Ref.~\cite{Fermilab} can be generalized from bilinears to 
four-fermion operators with Dirac structure $\Gamma_1\times\Gamma_2$. 
We can take one of the terms in the contraction of the 
lattice operator 
\begin{eqnarray} \label{eq:mix}
& \langle q(p'_q),b(p'_b)|\bar{q}\Gamma_1 b \bar{q}\Gamma_2 b |q(p_q),b(p_b)
\rangle_{\rm lat} = &\nonumber \\& N_q(p'_q)N_q(p_q)N_b(p'_b)N_b(p_b)
\bar{u}(p'_q)\Gamma_1 u^{\rm lat}_h(p'_b)\bar{u}(p_q)\Gamma_2 u^{lat}_h(p_b) 
+(\mathrm{additional} \, \mathrm{contractions}),&\nonumber\\ 
\end{eqnarray}
where $N_q(p)$ and $N_b(p)$ are normalization factors for the $q$ and $b$ 
one-particle states. Following the Fermilab interpretation in 
Ref.~\cite{Fermilab}, we demand that lattice and continuum amplitudes match 
through $\order(a\bm{p})$, 
\begin{eqnarray}
& {\cal Z}&\left( \bar{u}(\chi,\bm{p}'_q)
\Gamma_1 e^{-am_1^b/2}\left[1-\frac{i\bm{\gamma}\cdot \bm{p}'_ba} 
{2\sinh am_1^b}\right]u(\chi,0) 
\right.\nonumber\\&&\left.
\times \bar{u}(\chi,\bm{p}_q)\Gamma_2 e^{-am_1^b/2}
\left[1-\frac{i\bm{\gamma}\cdot \bm{p}_ba} {2\sinh am_1^b}\right]
u(\chi,0)\right)\,+\,a{\cal Z}D_nQ_n 
\nonumber\\ &=& \bar{u}(\chi,\bm{p}'_q)\Gamma_1 
\left[1-\frac {i\bm{\gamma}\cdot \bm{p}'_b} {2m_b}\right]u(\chi,0) \times 
\bar{u}(\chi,\bm{p}_q)\Gamma_2 \left[1-\frac 
{i\bm{\gamma}\cdot \bm{p}_b} {2m_b}\right]u(\chi,0)\nonumber\\
&&+O((\bm{p}a)^2)\,,
\end{eqnarray}
where $Q_n$ are dimension-seven lattice operators and $D_n$ their 
corresponding coefficients. These operators and coefficients 
are straightforward to identify ($m_b$ must be
identified with the Fermilab kinetic mass, $M_2$~\cite{Kronfeld00}).
There are two dimension seven operators contributing to the matching, 
$Q_1=\bar{q}\Gamma_1\bm{\gamma}\cdot \bm{D}b\bar{q}\Gamma_2 b$ 
and $Q_2=\bar{q}\Gamma_1 b \bar{q}\Gamma_2\bm{\gamma}\cdot \bm{D}b$, which  
will remove the $\bm{p}a$ discrepancy for appropriate values of the Wilson 
coefficients and the normalization constant. From the relation above, we 
find
\be
D_1=D_2=\left[\frac 1 {\sinh am_1^b}-\frac 1 {2am_2^b}\right]
\ee
and 
\be
{\cal Z}=e^{am_1^b}.
\ee  
Here, $m_1$ and $m_2$ are again the Fermilab rest and kinetic masses 
defined in \cite{Fermilab}.

However, through $\order(a\bm{p})$, adding these operators has the same effect
as inducing a ``rotation'' of the heavy field 
\be\label{eq:rotation}
b^r(x)=[1+a d_1 \bm{\gamma}\cdot \bm{D}]b(x)
\ee
where 
\be
d_1=D_1=D_2.
\ee   
This removes $\order(\Lambda_\mathrm{QCD}a)$ discretization errors 
in the operator. The leading errors are then $\order\left(
\left(\Lambda_{\mathrm{QCD}}a\right)^2\right)$ and $\order(\alpha_s
\Lambda_{\mathrm{QCD}}a)$. 
In the way we have set up the calculation, the open-meson propagators, 
Eq.~(\ref{eq:openmeson}), include the rotation.

\section{Matching of the Lattice Matrix Elements}

\label{sec:matching}

In order to cancel the scheme and scale dependence of
the Wilson coefficients in the effective Hamiltonian, we must relate the 
bare hadronic matrix elements of the lattice operators in Eq.~(\ref{susybasis}) 
to a continuum scheme. We perform that 
renormalization and matching perturbatively at one loop. In the lattice 
part of this renormalization calculation we use mean-field improved 
lattice perturbation theory~\cite{tadpoleimprv} to improve 
the convergence of the theory by resumming the tadpole contributions. 

Already at one loop, even in the continuum, the operators in 
Eq.~(\ref{susybasis}) mix with each other under renormalization. 
To extract the renormalized value of 
$\langle{\cal O}_1^q\rangle$, we use the following matching relation
\ba\label{matching}
\langle {\cal O}^{q}_1 \rangle ^{{\rm renor}}(\mu) = && 
\,{\cal C}\,\left\lbrace[ 1 + \alpha_s \cdot \zeta_{11}(\mu,m_b,am_b) ] 
\langle {\cal O}^{q}_1 \rangle^{\rm lat}  + \alpha_s \cdot \zeta_{12}(\mu,m_b,am_b)
\langle {\cal O}_2^{q} \rangle^{\rm lat} \right\rbrace\nonumber\\
&&+\order\left(\alpha_s^2,\alpha_s\Lambda_\mathrm{QCD}a\right)\, ,
\ea
where the renormalization coefficients $\zeta_{ij}$ are the difference between 
the renormalizations in the continuum and on the lattice, 
$\zeta_{ij}=Z_{ij}^{{\rm cont}}-Z_{ij}^{{\rm lat}}$. The continuum  
renormalization scale at which we perform the matching is $\mu$. The lattice 
spacing is $a$ and ${\mathcal C}$ is a factor which absorbs the lattice 
field normalization conventions. The values of $Z_{ij}^{{\rm cont}}$ are listed in  
Ref.~\cite{perturbativeHPQCD} and a detailed description of the calculation 
of the lattice renormalization coefficients will be given in 
Ref.~\cite{B0perturbative}. Table~\ref{tab:rhos} lists the tadpole-improved 
renormalization coefficients relevant  for the lattice data analyzed in this paper. 
For each lattice spacing and $b$ quark mass we show the  infrared (IR) finite part of 
the $Z_{ij}^{{\rm lat}}$'s as well as the corresponding $\zeta_{ij}$ (in the 
$\overline{{\rm MS}}$-NDR continuum scheme). 
The $\zeta_{ij}$ are IR finite since the IR divergent contributions to  
$Z_{ij}^{{\rm lat}}$ and $Z_{ij}^{{\rm cont}}$ cancel in the difference. 
All the coefficients $Z_{ij}^{{\rm lat}}$ in Table~\ref{tab:rhos} are 
between $0.3$ and $1$, which indicates a sensible behavior of the lattice 
perturbation series.

\begin{table}[t]
\caption{Values of the finite part of the lattice one-loop renormalization
coefficients $Z^{{\rm lat}}_{ij}$, the difference of the continuum and lattice
one-loop coefficients $\zeta_{ij}$ needed in (\ref{matching}) for the $0.12~{\rm fm}$ 
and
$0.09~{\rm fm}$ lattices, and the coupling $\alpha_s$ used in the matching relation.
The continuum ($\overline{{\rm MS}}$-NDR) scale used in the matching is $\mu=m_b$.
\label{tab:rhos}}
\begin{tabular}{c|c|cccc|c}
\hline\hline

\vspace*{0.2cm}

$\approx a~({\rm fm})$ & $am_b$ & $Z_{11}^{{\rm lat, \,finite}}$ & 
$Z_{12}^{{\rm lat,\, finite}}$  & $\zeta_{11}^{\overline{{\rm MS}}{\rm -NDR}}$ 
& $\zeta_{12}^{\overline{MS}-NDR}$ & $\alpha_s$ \\
\hline
0.12 & 2.1881  & -0.726  & -0.325  & 0.1998 & -0.312 & 0.32 \\
0.09 & 1.7728  & -0.945  & -0.369  & 0.3041 & -0.268 & 0.26 \\
\hline
\hline
\end{tabular}
\end{table}

In order to apply the matching relation Eq.~(\ref{matching}), we need to choose 
both a scale $\mu$ and a value for the strong coupling constant $\alpha_s$. 
For the scale, we use the bottom quark mass and, in that way, eliminate  
higher-order logarithmic contributions that come in powers of 
$\log(\mu/m_b)$. For the strong coupling constant, we use the 
renormalized coupling in the $V$-scheme~\cite{lepage-1993-48} evaluated at  
a scale $q^*$, as in Ref.~\cite{Quentin2005}. The scale $q^*$ should be the size 
of a typical gluon loop momentum in
this process and can be calculated using the methods outlined in
Refs.~\cite{lepage-1993-48,BLM83}. Here, we use $q^*=2/a$ which is close to 
the calculated value for heavy-light currents using the same actions we are
employing~\cite{lepage-1993-48,HLM03}. This is
justified since the contributions coming from the current renormalization
are larger than the intrinsic four-quark contributions~\cite{B0perturbative}.
The values of $\alpha_V$ are determined from the static-quark potential in a 
manner similar to that described in Ref.~\cite{Quentin2005} and are also given in 
Table~\ref{tab:rhos}.

\section{Fitting Method and Statistical Errors}

\label{sec:fitting}

The correlation functions are calculated at four different time sources $t$, and 
then averaged over time sources. For the $a\approx 0.12~{\rm fm}$ ensembles, 
$t_0=0,16,32,48$, and for the $a\approx 0.09~{\rm fm}$ ensembles, $t_0=0,24,48,72$. 
The statistical errors in the data and fits decrease with each additional time source  
by approximately what is expected, 
suggesting that the correlators from different time sources  
are weakly correlated and statistical power is gained by averaging.

In order to extract the renormalized matrix elements, we tried two 
methods for the correlator fits. In the first method, we fit the bare 
correlators and combine the results afterwards with the matching 
coefficients in Sec.~\ref{sec:matching} to get the renormalized 
matrix elements. In the second method, we first apply the matching coefficients
to the correlators for each configuration and then perform the fits to obtain the 
renormalized matrix elements. The central values are 
nearly identical with both methods, but the errors are slightly better and the 
fits more stable with the latter, so for the rest of this article we  discuss  
only the results obtained with the second method.

\subsection{Description of the Fitting Method and Stability Tests}

\label{fitting}

The heavy-quark in the two-point and three-point correlation functions is 
always rotated at the source as explained in Sec.~\ref{sec:openmeson}. For 
three-point functions, we smear the heavy quarks at the sink using a function 
based on the quarkonium 1S wavefunction~\cite{Richardson1S,DiPierro2002}. 
For two-point functions, at the sink we either rotate local heavy quarks, 
or we smear them with a 1S wavefunction. 
Smearing greatly improves the overlap with the ground state.
The additional rotation at the sink is to ensure that the local-local meson
correlator is positive-definite. The naive light-quark propagator is always
local at source and sink.

The two-point and three-point correlators used to determine the matrix 
element $\langle{\cal O}_1^q\rangle$ on a particular ensemble and for a particular 
choice of valence-mass $m_q$ are fit simultaneously using the Bayesian fitting 
approach described in Refs.~\cite{Lep2001,Morningstar01}. 
For the matrix elements on the coarse ensembles,  
we find the smallest errors and greatest stability using three 
correlators (two two-point correlation functions and one three-point correlation 
function):
   \begin{itemize}
   \item[] - $C_{PS}$ in Eq.~(\ref{eqn:CPS}) with local source and local sink.
   \item[] - $C_{PS}$ with local source and 1S smeared sink.
   \item[] - $C_{{\cal O}_1^{q}}$ in Eq.~(\ref{eqn:CQ}) with local source and 1S 
smeared sink.
   \end{itemize}
For the fine ensembles the best results are obtained with only one two point 
function and one three-point function:
    \begin{itemize}
   \item[] - $C_{PS}$ with 1S smeared source and 1S smeared sink.
   \item[] - $C_{{\cal O}_1^{q}}$ with local source and 1S smeared sink.
   \end{itemize}

The prior central values function as the initial starting guesses for our fits. Hence we 
choose ground-state values guided by our data to help the fits converge. The prior central 
values for the ground-state masses are obtained from effective mass plots. For the overlap 
factors $Z^d_0$ and $Z^{1S}_0$, where superscripts $d$ and 
$1S$  denote factors corresponding to local or 1S smeared sources/sinks, we examine the 
amplitude of the $B$  meson propagator with the exponential of the ground state removed. 
We do the same for $O_{00}$, where the $Z^{1S}_0$ amplitudes are accounted for. 
The prior widths are taken to be large compared 
with the statistical error of the parameters as reported by the fitter  
to avoid influencing our fit results by our choice of priors. 
For the higher states' overlap factors, the prior width is chosen based on the expectation 
that the overlaps should not be larger than the corresponding ground state ones.        
The energy differences have prior central values and widths that allow them to vary
from $\Delta E_{i+1,i}\equiv a(E_{i+1}-E_i) \approx 0.14-0.37$, where experimental 
values~\cite{PDG2011} have been used as a guide. We checked that the prior widths for 
all fitting parameters are large enough so they do not influence the central value of 
relevant quantities extracted from the fits.

The same priors are used for all ensembles, 
except for the masses of the regular and oscillating ground states, $E_0$ 
and $E'_0$ respectively. 
These parameters are strongly determined by 
the data, and very different at each lattice spacing, therefore 
the prior choice must also be lattice-spacing dependent. 
Appendix~\ref{app:priors} contains a list of the prior central values and widths 
we use in the calculation.

\begin{figure}[t]
\includegraphics[angle=-90,width=0.60\textwidth]{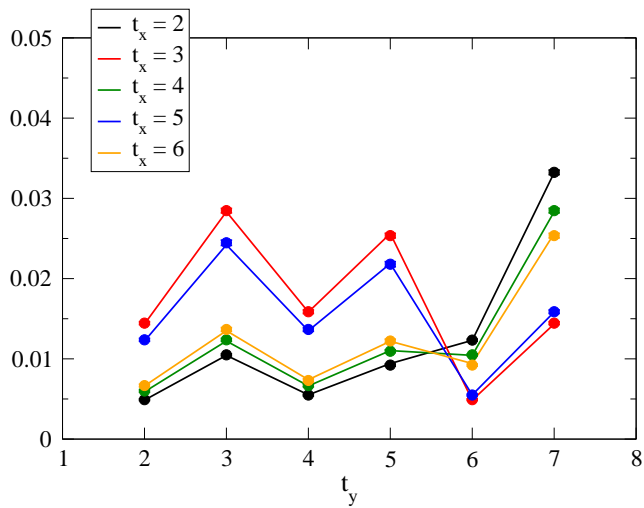}
\caption{\label{fig:correlatorfit} Comparison of  a fit with correlator data for the 
correlation function $C_{{\cal O}_1^q}(t_x,t_y)$ at  
fixed $t_x=2$-$6$ (labeled by color) and as a function of $t_y$, for the 
$a\approx0.09~{\rm fm}$ ensemble with quark masses $0.0124/0.031$ and valence 
quark mass $am_q=0.031$. The lines connect the fit function results for integer 
values of $t_x,t_y$ coming from the same  
single fit and evaluated at specific $t_x$, and the dots are the average over 
simulation data. Statistical errors 
on the simulation data are smaller than the plot symbols. The fit results describe 
very well the oscillation in time shown by the data.}
\end{figure}

\begin{figure}[t]
\includegraphics[angle=-90,width=0.60\textwidth]{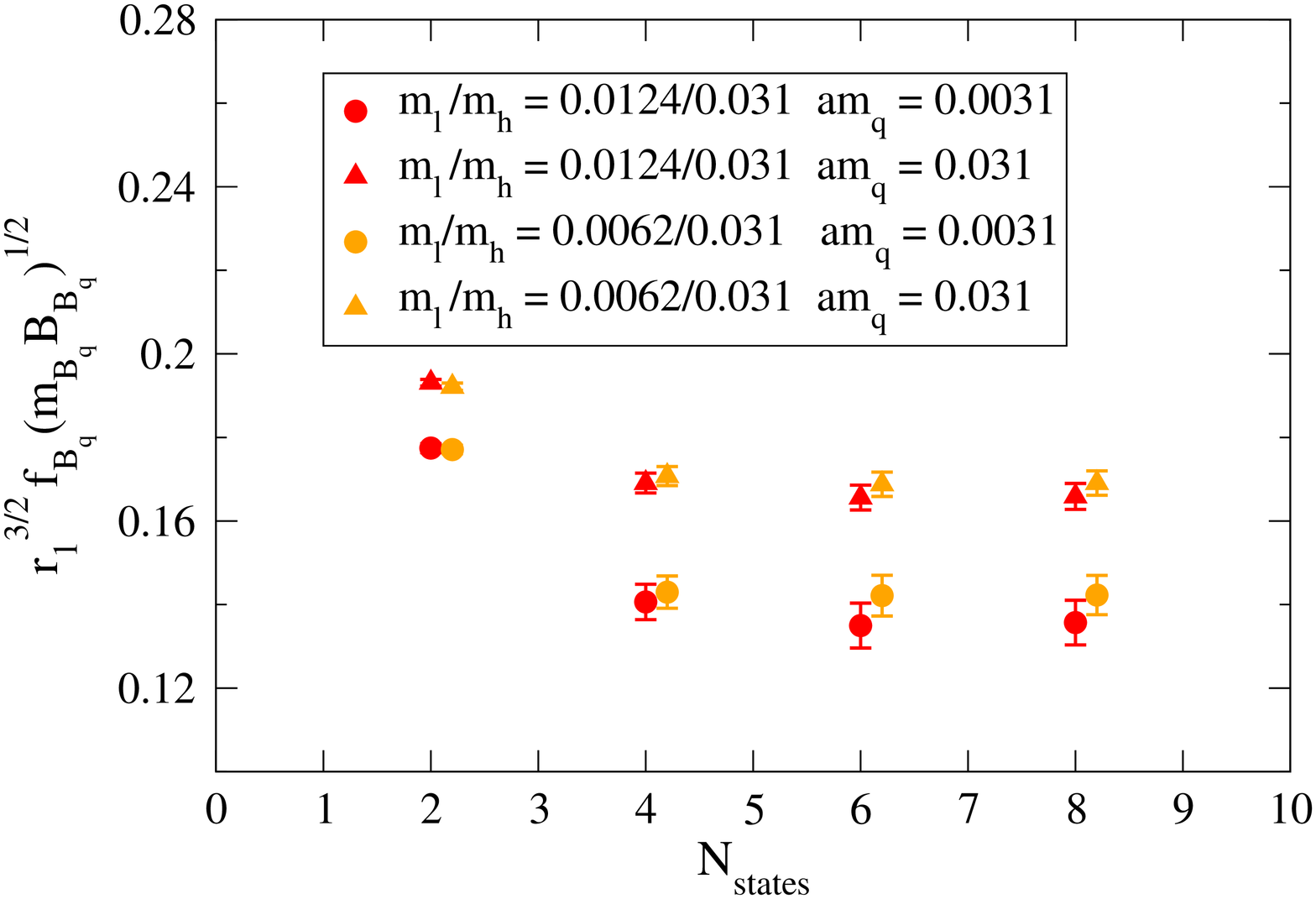}
\caption{\label{fig:fbB_nstates} $r_1^{3/2}f_{B_q}\sqrt{M_{B_q}B_{B_q}}$ for the two 
$a\approx0.09~{\rm fm}$ ensembles for $am_q=0.031,0.0031$ as a function of the 
number of states $N_{{\rm states}}$. The fit results for $f_{B_q}\sqrt{M_{B_q}B_{B_q}}$ 
reach a plateau for $N_{{\rm states}}\ge4$.}
\end{figure}

Statistical errors are estimated with the bootstrap method. Specifically, 
for each ensemble and valence mass, 500 bootstrap ensembles are constructed 
from the original ensemble by sampling with replacement.  A fit is then 
performed to each ensemble. We find that, as long as the bootstrap 
ensembles are larger than $\sim 100$, the estimated error is independent of 
bootstrap ensemble size. For fitting methodology checks and plotting purposes in 
Figs.~\ref{fig:correlatorfit}--\ref{fig:xi_NNLO_cf}, statistical 
errors in the parameters are
estimated by the average 68\% bootstrap error, which is defined
as half of the distance between the two points at which 16\% of the 
distribution has a higher (lower) value.

Autocorrelations necessarily exist between correlation functions calculated 
on different configurations within an ensemble and can be minimized by 
binning the data.  
The autocorrelations are observable only in a few ensemble and valence masses. 
In many ensembles and mass combinations, the noise is large enough that the 
autocorrelations are not observable. We choose a conservative bin size of 4.

The number of states included in the sum in Eq.~(\ref{eqn:CQ}) and 
the time ranges we use in the fits are shown in Table~\ref{tab:nt}. 
The minimum time slice is fixed to be the
same for all the correlators in the fit, three-point as well as two-point. However, 
the maximum time is fixed separately for the two- and three-point functions. 
Following Ref.~\cite{Lep2001}, the number of states are determined by first 
performing the fit using 1+1 states (1 regular state + 1 oscillatory state) 
starting at large time slices, where the higher energy states no longer 
contribute significantly and a good $\chi^2$ per degree of freedom (d.o.f.) is 
obtained ($\approx 1$).  The fit is then performed using one lower time slice as 
the starting time $t_{\text{min}}$, and this is repeated, reducing $t_{\text{min}}$ until 
the $\chi^2/{\rm d.o.f.}$ is no longer reasonable, $\gtrsim 1.5$.  Then an 
additional pair of states is added to the model function, and the process iterated. 
Once the timeslice $t=2$ can be included, that number of states is used in our  
central-value fits.\footnote{Time slices $t=0$ and $t=1$ contain unconstrained 
contamination from higher energy states. At $t=0$ all states contribute because they 
have the same exponential weighting, and  $t=1$ is contaminated by higher energy 
states because  the degrees of freedom of staggered fermions spread over two time slices.}

  \begin{table}[t]
    \centering
 \caption{\label{tab:nt} Number of states and time ranges used for each
correlator in the fits for both the $a\approx 0.12~{\rm fm}$ and 
$a\approx 0.09~{\rm fm}$ ensembles. For the number of states, the first value
indicates the number of regular states and the second one the number
of oscillating states. The labels between parentheses in the first column indicate
the type of source/sink in that correlator.}
  \begin{tabular}{ccc}
 \hline
 \hline
 Correlator & Number of States & Time Range \\
\hline
 $C_{PS}$ (local/local) & 3+3 & 2--20\\
 $C_{PS}$ (local/1S) & 3+3 & 2--20\\
 $C_{Q^i_{q}}$ (local/1S) & 2+2 & $(2-10)\times(2-10)$ \\
 \hline
  \hline 
  \end{tabular}
  \end{table}

For the three-point function, we fit using $N_{{\rm states}}
=N_{{\rm states}}^{{\rm regular}}+N_{{\rm states}}^{{\rm oscillating}}=2+2=4$ and  
timeslices $t_x,t_y\in[2,10]$ for all ensembles.  The two-point functions are 
fit for $t\in[2,20]$ using $3+3$ states. The output of these fits successfully  
describe the oscillations in the correlation functions as can be seen in 
Fig.~\ref{fig:correlatorfit}, which shows the typical behavior in one 
of the ensembles analyzed.  

In order to check that this number of states is sufficient, 
we add more states and examine the stability of the fits. 
Stability plots over numbers of states for the $a\approx0.09~{\rm fm}$ ensembles,  
which illustrate the typical behavior of our fits, are shown in 
Fig.~\ref{fig:fbB_nstates}. The stability of central values and errors is very 
good for $N_{{\rm states}}\ge 4$ in all cases.

\section{Chiral Perturbation Theory}

\label{sec:chiralfits}

The light sea- and valence-quark masses that are used in our lattice 
simulations have unphysically large values, with our lightest pion mass  
$\approx240$ MeV. To obtain information about the quark-mass dependence of the 
relevant matrix elements, which allows us to extrapolate our results to the 
physical masses, we perform our calculation at six sea $\times$ six 
valence quark masses, 
thus including numerous partially quenched data points. In addition, the leading-order 
taste violations, which arise at $\order(a^2\alpha_s)$,  are included 
in the theory and then removed when the extrapolation is performed using  
rooted heavy-meson staggered chiral perturbation theory 
(rHMS$\chi$PT)~\cite{SharpeandLee,aubin-2003,aubin-2006-73}. 
The $\chi$PT expression 
for $\langle{\cal O}_1^q\rangle$, as well as for the matrix elements 
of all the other operators in the $\Delta B=2$ effective Hamiltonian 
was first described in Ref.~\cite{detmold-2006} for partially quenched 
Wilson-type quarks in the framework of continuum heavy-meson
 $\chi$PT (HM$\chi$PT). With staggered fermions, we also must include the 
effects of taste-violating interactions, using rHMS$\chi$PT~\cite{aubin-2006-73}. 

With the four-quark operators, a careful examination of the Fierz properties 
shows that there are additional operators with both wrong taste and 
\emph{spin}, {\it i.e}, wrong $(\Gamma_1,\Gamma_2)$. As far as we know, 
this property of local heavy-staggered four-quark operators has not been 
discussed in the literature before. The needed rHMS$\chi$PT expressions are 
derived in Ref.~\cite{B0sCHPT}. We became aware of these contributions after 
our analysis was nearly complete, so we have not included them in the chiral 
fit functions used here. We do, however, estimate the associated systematic 
error on $\xi$ in our error budget (cf. Sec.~\ref{sec:wrongspin}). 
Explicit expressions for the chiral fit functions used in this work are 
given in Appendix~\ref{app:SCHPT}.

The NLO rHMS$\chi$PT in Eq.~(\ref{eq:ChPTO1}) 
and subsequent equations in the appendix can be schematically written as 
\ba
\label{eq:chipt0}
&&\left\langle\bar{B}_q|{\cal O}_1^q|B_q\right\rangle=\frac 2 3 
M_{B_q}^2 f_{B_q}^2 {B_q}=
\nonumber\\&&  M_{B_q} 
\alpha \left[1+({\cal W}_q+{\cal T}_q+{\cal Q}_q)+L_vm_q+L_s(2m_l+m_h)
+L_aa^2\right]\, ,
\ea
where $\alpha$, $L_v$, $L_s$,  and $L_a$ are low-energy 
constants (LECs) to be determined from fitting the data,  
the factor of $M_{B_q}$ comes from the HQET normalization of states,
and the masses $m_q$, $m_l$, and $m_h$ are the light valence, 
light sea, and strange sea-quark masses, respectively. 
The light sea quarks are treated as degenerate, and the isospin average 
is used, {\it i.e.}, $\hat m=(m_u+m_d)/2$.  For staggered quarks the 
taste-nonsinglet pseudoscalar meson masses are split
\be\label{eq:Mij}
M^2_{ij,\rho}=\mu (m_i + m_j)+a^2 \Delta_{\rho}\, , 
\ee
where $m_i$ and $m_j$ are the quark masses and the sixteen meson masses are 
labeled by their taste representation, $\rho=P,A,T,V,I$.  The parameters 
$\mu$ and the $\Delta_{\rho}$'s are determined from lattice calculations for 
pions and kaons \cite{aubin-2004-70}. Their values are collected in 
Table~\ref{tab:ChPTpar}. 

The chiral logarithms, ${\cal W}_q,\, {\cal T}_q,\, {\cal Q}_q$, stem from 
wavefunction renormalization, tadpole, and sunset diagrams,  
respectively. The explicit expressions can be found in Appendix~\ref{app:SCHPT}.  

In this work, we do not include the effects of the hyperfine splitting $\Delta^*$ 
or the light flavor splittings $\delta_k$ defined in the Appendix. 
The wrong-spin terms contribute to the tadpole and sunset diagrams~\cite{B0sCHPT}.

When extrapolating to the physical point, we set 
the parameters $\Delta_{\rho}$ and $\delta'_{A,V}$ (which describe discretization 
effects) and the lattice spacing to zero, and set the sea quark masses 
to their physical values, $m_l\to(m_u+m_d)/2$, and $m_h\to m_s$.  We then obtain 
$\langle \bar{B}_d^0 |{\cal O}_1^d|B_d^0\rangle$ or $\langle \bar{B}_s^0|{\cal O}_1^s|
B^0_s\rangle$ by setting $m_q= m_d\, \mathrm{or}\, m_s$.  Thus, it is an 
extrapolation to the $u$- and $d$-quark masses and an interpolation to the 
$s$-quark mass.  
      
An additional consideration is that $SU(3)$ NLO $\chi$PT may not be 
valid for data with masses as large as the strange quark's.  
It would be desirable to include NNLO contributions to test the validity of 
the NLO expression, but the effort needed to calculate the NNLO logs is 
prohibitive. It is reasonable instead to test the chiral expansion by 
including just NNLO analytic contributions. Wherever the quark masses or 
splittings are large enough for such analytic NNLO terms to be significant, 
the NNLO logarithms should be slowly varying and well approximated
by the analytic terms. We follow this strategy and supplement the NLO 
rHMS$\chi$PT expressions with NNLO analytic terms in our fits, with prior 
constraints estimated based on $\chi$PT power counting as explained
in the next section.

\subsection{Parametrization of the Chiral Expression}

\label{sec:parmCHPT}

Dimensionful quantities are extracted first in units of the lattice spacing and 
then converted to their physical values using the $r_1$ scale~\cite{r11,r12}. 
This absolute scale is defined as $r_1^2F(r_1)=1.0$, 
where $F(r_1)$ is the force between static quarks. 
In our chiral fits, all parameters are first converted to units of 
$r_1$ by multiplying by the relative scale $r_1/a$. The  values for $r_1/a$ on
every MILC ensemble used in our calculation are listed in Table~\ref{tab:ensemblesa}. 
After the chiral-continuum extrapolation, we convert from $r_1$ units with a physical 
value of $r_1$. We take  
the result obtained by combining the 2009 MILC determination of 
$r_1f_\pi$~\cite{fpiMILC09} and the PDG value of $f_\pi$~\cite{PDG2011}. 
Following Ref.~\cite{decayconstants2011}, the 
error for $r_1$ is determined by averaging the MILC value with the HPQCD value 
in \cite{r1HPQCD} and then by inflating the uncertainty to take conservatively 
into account the possible correlations coming from the use of the same 
configurations in both determinations. The final value we use is $r_1=0.3117(22)$ 
\cite{decayconstants2011}. The error associated with $r_1$ has a very small effect 
on the dimensionless quantity $\xi$.

The dominant lattice artifacts to take into account in our 
rHMS$\chi$PT expressions are expected to be taste violating contributions 
of $\order(a^2\alpha_s^2)$, since the $\order(a^2\alpha_s)$ taste-violating 
effects are absent for asqtad quarks. We parametrize 
these effects in our fits by defining a quantity $A^2_a$, which is the ratio 
of the size of taste violations on lattices with spacing $a$ to those on the 
$a\approx 0.12$ fm lattices. Thus $A^2_{0.12\; {\rm fm}}=1$ and
\ba
&&A_{0.09~{\rm fm}}^2\equiv\frac {(\alpha_s^2a^2)_{0.09~{\rm fm}}} 
{(\alpha_s^2a^2)_{0.12~{\rm fm}}}\sim0.35\, .
\ea
The NLO rHMS$\chi$PT function in Eq.~(\ref{eq:chipt0}) can then be rewritten as
\ba
\label{eq:chipt2}
\beta_q \equiv &&\sqrt{\frac 3 2 \langle\bar{B}_q|{\cal O}_1^q|B_q\rangle/{M_{B_q}}}=
f_{B_q}\sqrt{M_{B_q}B_{B_q}}\nonumber\\
 && = \beta^{\chi} \left[1+\frac 1 2 ({\cal Q}_q+{\cal W}_q+{\cal T}_q)
+\frac{L_v}{2}M^2_{qq}+\frac{L_s}{2}(2M^2_{ll}+M^2_{hh})+\frac{L_a}{2}
A_a^2\right]\,,
\ea
where $\beta^\chi =\sqrt{\alpha}$. 
The masses $M_{ij}$ are defined in Eq.~(\ref{eq:Mij}), but here we disregard 
$a^2$ corrections in the masses since they can be absorbed by a redefinition of 
the low energy constants at higher order in the chiral expansion. 
To the NLO expression above we add, inside the square brackets, the allowed 
NNLO analytic terms, which contribute with seven more unknown LECs
\ba \label{eq:NNLO}
&&Q_1M_{qq}^4 + Q_2(2M^2_{ll}+M^2_{hh})^2 + Q_3M^2_{qq}(2M^2_{ll}+M^2_{hh})\\ 
\nonumber &&+ Q_4(2M^4_{ll}+M^4_{hh}) + P_1A_a^2M^2_{qq} + P_2A_a^2(2M^2_{ll}
+M^2_{hh}) + P_3A_a^4\, .
\ea
In the above expressions, we have suppressed the factors of $r_1$ for 
simplicity. They can be deduced using dimensional considerations. In these 
expressions, which are the ones we use as fit functions, we write the 
analytic terms for convenience as functions of 
the pseudoscalar masses $M_{ij}$ rather than the quark masses.

The ratio $\xi$ can be extracted by first interpolating $\beta_q$ to $m_q=m_s$ and 
extrapolating to $m_q=m_d$ separately according to expressions (\ref{eq:chipt2}) and 
(\ref{eq:NNLO}),  and then forming the ratio $\beta_{s}/\beta_d$. Alternatively, one 
can consider the ratio of chiral expressions and expand up to NNLO to obtain
\ba \label{eq:xichipt}
\xi'=\frac {\beta_{q'}} {\beta_q}=\xi\frac{\sqrt{M_{B_{q'}}}}{\sqrt{M_{B_q}}} & = &
1+\frac{1}{2}({\cal Q}_{q'}+{\cal W}_{q'}+{\cal T}_{q'}
-{\cal Q}_q-{\cal W}_q-{\cal T}_q)+
\frac{L_v}{2}(M^2_{q'q'}-M^2_{qq})  \nonumber\\
&& + Q_1(M^4_{q'q'}-M^4_{qq}) + Q_3(M^2_{q'q'}-M^2_{qq})(2M^2_{ll}+M^2_{hh})
\nonumber\\
&& + P_1(M^2_{q'q'}-M^2_{qq})A_a^2\, ,
\ea
with $m_{q'}$ fixed to the value closest to $m_s$. In Eq.~(\ref{eq:xichipt}) 
we disregard the NNLO terms coming from squaring the NLO terms in the denominator, 
since they are not necessary to obtain good fits 
and they are difficult to disentangle from those already included. We can then 
interpolate/extrapolate to $m_{q'}=m_s$ and $m_q=m_d$. 
We call these two strategies for the chiral and continuum extrapolation of $\xi$ 
the indirect and direct methods, respectively. 
Many of the fit parameters cancel in the chiral expression for $\xi'$ 
in Eq.~(\ref{eq:xichipt}), improving the reliability and 
stability of the fits.  In addition, discretization errors of 
$\order(\alpha_s\Lambda_\mathrm{QCD}a,(\Lambda_\mathrm{QCD}a)^2)$ from 
the heavy-quark action that are not included in the chiral perturbation 
theory, partially cancel in the ratio. We thus choose this method as our 
preferred fitting strategy.

  \begin{table}[t]
    \centering
\caption{\label{tab:ChPTpar} Inputs for the priors of the free parameters 
and for the fixed parameters 
in the fits. The NLO low energy constants $L_v$, $L_s$, and $L_a$ are not 
constrained in the fits.
The parameter $s$ is given by the quantity $1/(8\pi^2(r_1f_\pi)^2)$. We do 
not consider errors on the slope $\mu r_1$ or the taste splittings 
$r_1^2a^2\Delta_\rho$ because those have negligible effect on the final results. 
In the right hand side table, the two last columns correspond to lattice spacings 
$a\approx 0.12~{\rm fm}$ and $a\approx 0.09~{\rm fm}$. 
See the text for explanations of the choices of parameters.}

\hspace*{0.2cm}
  \begin{tabular}{ccc}
\hline 
\hline
\multicolumn{3}{c}{Fit parameters (central value$\pm$width)}\\
\hline
  & $a\approx 0.12~{\rm fm}$ & ($a\approx 0.09~{\rm fm}$) \\ 
\hline 
$\beta$ &  \multicolumn{2}{c}{$1\pm 1$} \\
$g_{B^*B\pi}$ & \multicolumn{2}{c}{$0.51\pm 0.20$}\\
\hline
$r_1^2a^2\delta'_V$ & $0.0\pm0.07$ & $(0.0\pm0.07)\times 0.35$\\
$r_1^2a^2\delta'_A$ & $-0.28\pm0.06$  & $(-0.28\pm0.06)\times 0.35$ \\
\hline
$L_v$ & \multicolumn{2}{c}{unconstrained} \\
$L_s$ & \multicolumn{2}{c}{unconstrained} \\
$L_a$ &   \multicolumn{2}{c}{unconstrained} \\
$Q_{1-4}$ & \multicolumn{2}{c}{$0\pm s^2$} \\
$P_{1-3}$ & \multicolumn{2}{c}{$0\pm s^2$} \\
\hline
\hline
\end{tabular}
\hspace*{0.5cm}\begin{tabular}{ccc}
\hline\hline
\multicolumn{3}{c}{Input (fixed) parameters}\\
\hline
\hline
 & \hspace*{-0.3cm}$a\approx 0.12~{\rm fm}$ & $a\approx 0.09~{\rm fm}$ \\
\hline 
$f_{\pi}r_1$ & \multicolumn{2}{c}{0.2106}\\ 
\hline
$\mu r_1$ & 6.234 & 6.382  \\
\hline
$r_1^2a^2\Delta_P$ & 0 & 0 \\
$r_1^2a^2\Delta_V$ & 0.439 & 0.152 \\
$r_1^2a^2\Delta_T$ & 0.327 & 0.115  \\
$r_1^2a^2\Delta_A $& 0.205 & 0.0706  \\
$r_1^2a^2\Delta_I$  & 0.537 & 0.206 \\
\hline
\hline 
\end{tabular}
\end{table}

\subsection{Results from the chiral fits}

In order to perform the chiral fits, we first create 200 bootstrap samples of 
$\beta_q$ 
for each sea- and valence-quark mass combination from the two- and three-point 
correlator fits.  The bootstrap data is then fit to the chiral expression 
using Bayesian techniques. The fits are simultaneously performed to all 
ensembles in Table~\ref{tab:ensemblesa}.  

The input and fit parameters are set as in Table~\ref{tab:ChPTpar}.  
We do not impose any constraint on the NLO low energy constants. 
For the NNLO LECs we use prior widths 
based on a simple power counting argument. The NLO 
analytic terms should be of magnitude similar to the NLO logs, which 
are $\sim m_\pi^2/(8\pi^2f_{\pi}^2)=s(r_1m_\pi)^2$, (with $s \equiv 1/(8 
\pi^2(r_1 f_\pi)^2)$). Hence, the NNLO terms are~$\sim \left(m_\pi^2/
(8\pi^2f_{\pi}^2)\right)^2=s^2(r_1m_\pi)^4$. The taste-violating hairpin
parameters, $\delta_V'$ and $\delta_A'$, were also determined from lattice 
calculations for pions and kaons in Ref.~\cite{MILCasqtad}. 
We constrain the parameters $\delta_V'$ and $\delta_A'$  in our fits using the
results of Ref.~\cite{MILCasqtad} as prior central values and widths. 
We also take the effective coupling of the $B^* B \pi$ 
interaction, $g_{B^*B\pi}$, as a fit parameter in our analysis. 
The prior central value and width we use for this parameter, shown in 
Table~\ref{tab:ChPTpar}, covers the main ranges  of determinations of 
$g_{B^*B\pi}$ 
\cite{Stewart98,CDetal97,CLEOgpi02,ABBetal02,AGRS05,OMO08,ALPHAgpi10}, 
as discussed in Ref.~\cite{JackBtoD*1}. A more recent, precise value of 
$g_{B^*B\pi}$, obtained with $N_f=2+1$ domain wall fermions and static $b$ 
quarks~\cite{DLMgbbspi}, was not yet available when this stage of the 
analysis was carried out. Nevertheless, the result obtained by the authors in 
Ref.~\cite{DLMgbbspi},  $0.449 \pm 0.047\pm0.019$, falls well within 
the prior central value and width considered here.
For the pion decay constant, we use the PDG value, 
$f_\pi=(130.41\pm0.20)~{\rm MeV}$~\cite{PDG2011}.

The fit results for $\xi$ using different ansatzes 
for the fitting function and the direct and indirect methods explained in 
Sec.~\ref{sec:parmCHPT} are listed in Table~\ref{tab:directfit_cf}. The results and 
errors obtained using the direct and indirect methods agree very 
well, especially when NNLO terms are included. This constitutes a good check 
of how well our results are encompassing higher-order terms in the chiral expansion, 
which are different in these two methods. 

\begin{table}[t]
    \centering
\caption{\label{tab:directfit_cf} Results from the rHMS$\chi$PT fits. Errors 
are only statistical and obtained from 200 bootstrap samples. For a full 
discussion of systematic errors, see Sec.~\ref{sec:Errors}.}
  \begin{tabular}{ccccc}
\hline
\hline
  Ansatz & $\chi^2/{\rm d.o.f.}$ & $\xi$ Direct &  $\chi^2/{\rm d.o.f.}$ & $\xi$ Indirect\\
\hline
NNLO & 0.45 &   $1.268^{+0.035}_{-0.044}$  & 0.23 &
$1.255^{+0.034}_{-0.041}$ \\
NLO    & 0.78 &   $1.284^{+0.018}_{-0.016}$  & 0.49 & $1.262^{+0.008}_{-0.012}$ \\
\hline\hline
\end{tabular}
\end{table}

\begin{figure}[t]
\includegraphics[angle=-90,scale=0.4]{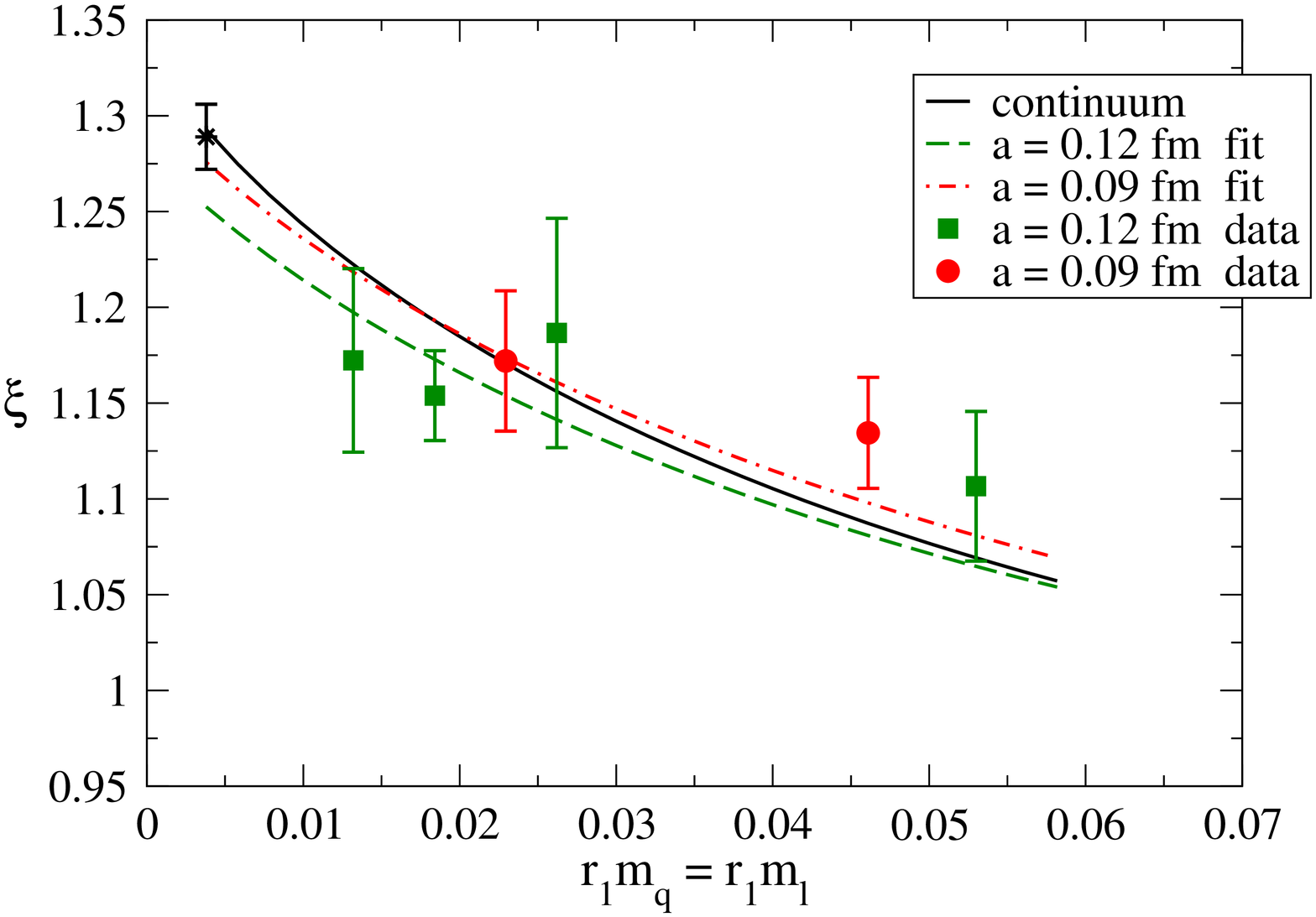}\\
\vspace*{-0.2cm}
\includegraphics[angle=-90,scale=0.4]{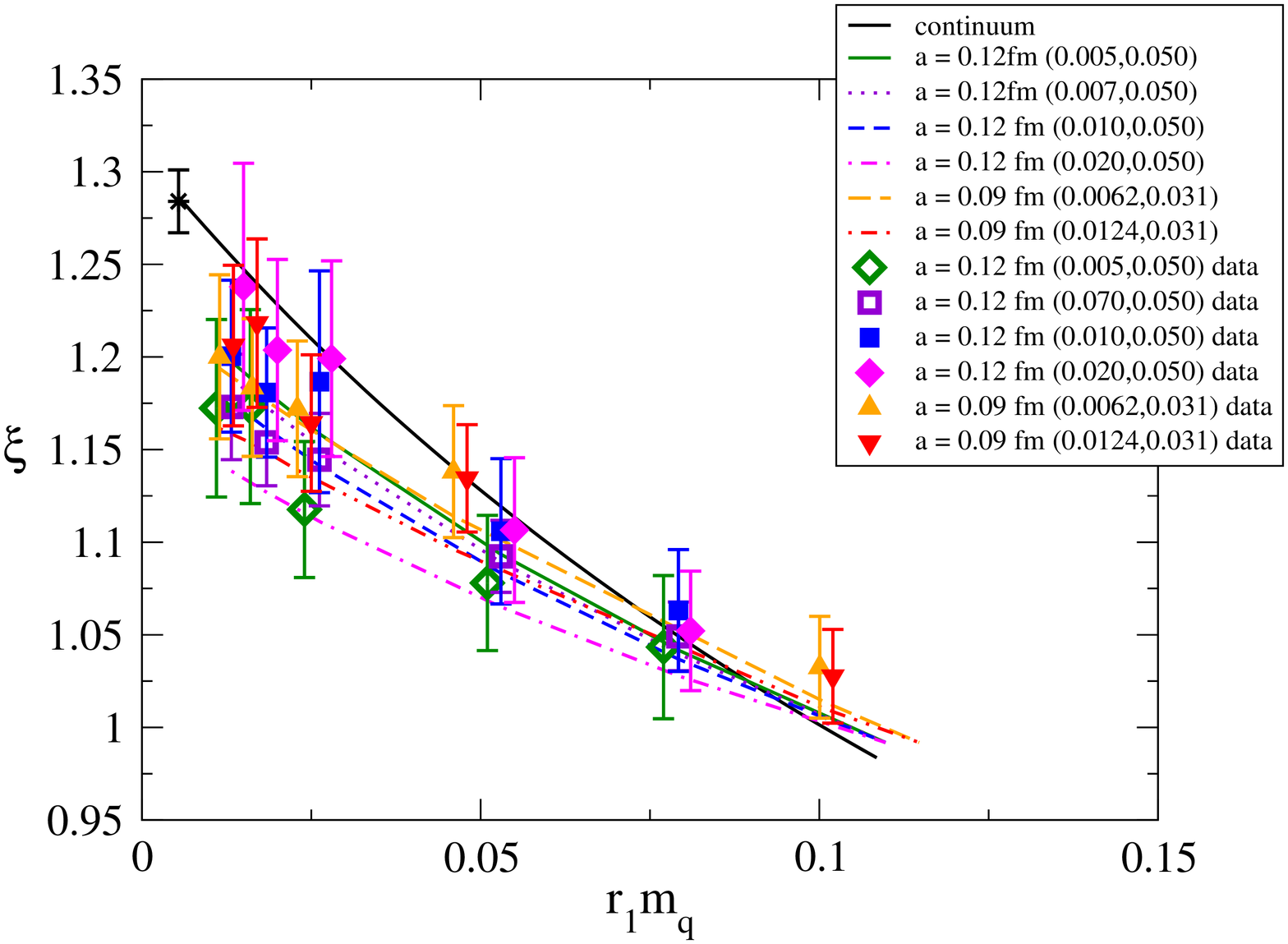}
\caption{\label{fig:xi_NLO_cf}  Fits results using the NLO rHMS$\chi$PT, 
first line in Eq.~(\ref{eq:xichipt}). The (black) star is the physical 
value of $\xi$ in both plots. The top plot shows only the full QCD data, while 
the bottom one shows all the data included in the fits. The (green) squares and 
(red) circles and lines in the upper plot represent the $0.12~{\rm fm}$  
and $0.09~{\rm fm}$ data and fit results respectively. 
In the bottom plot, each color (symbol) labels a different ensemble.
}
\end{figure}

\begin{figure}[t]
\includegraphics[angle=-90,scale=0.4]{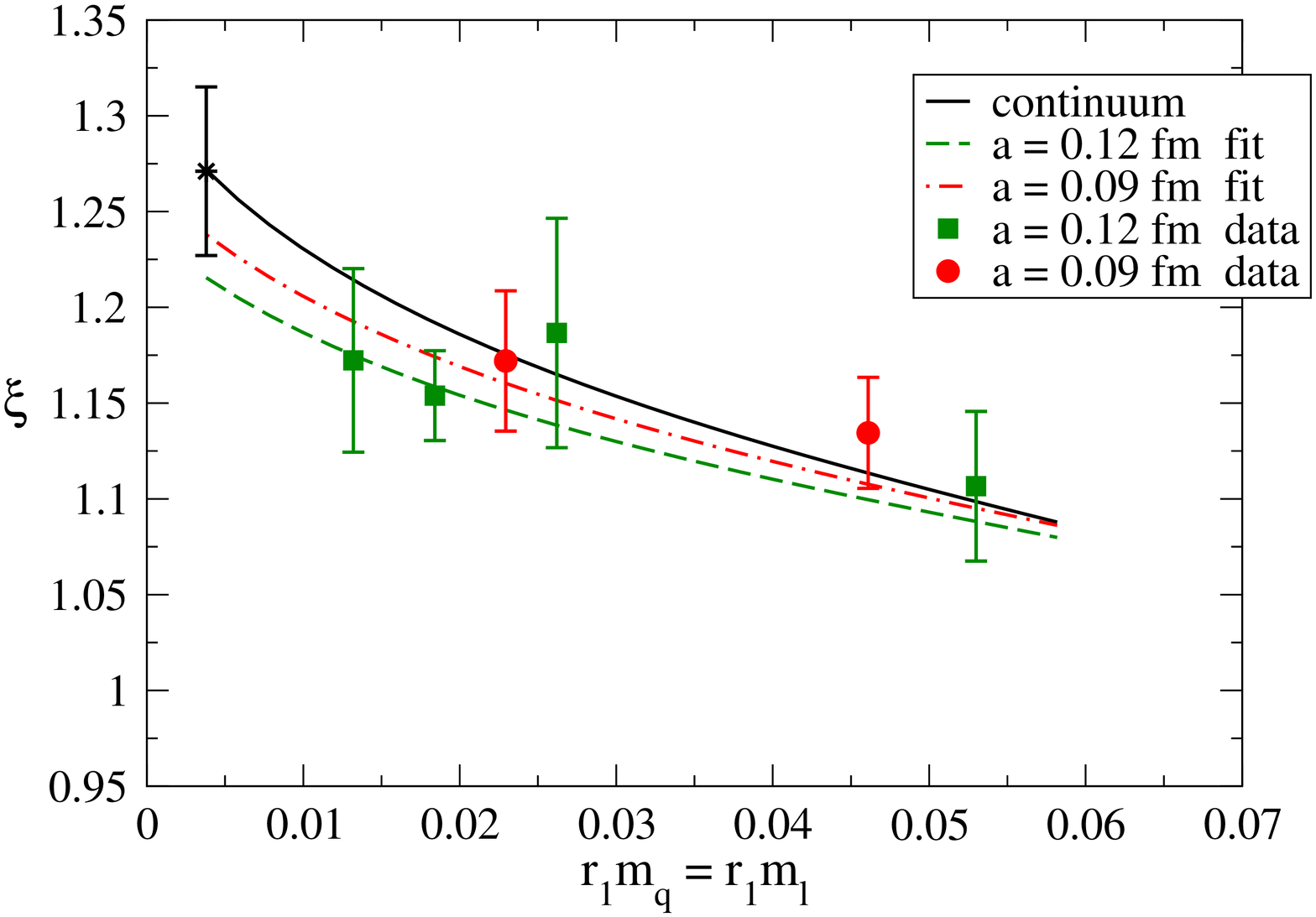}\\

\vspace*{-0.2cm}
\includegraphics[angle=-90,scale=0.4]{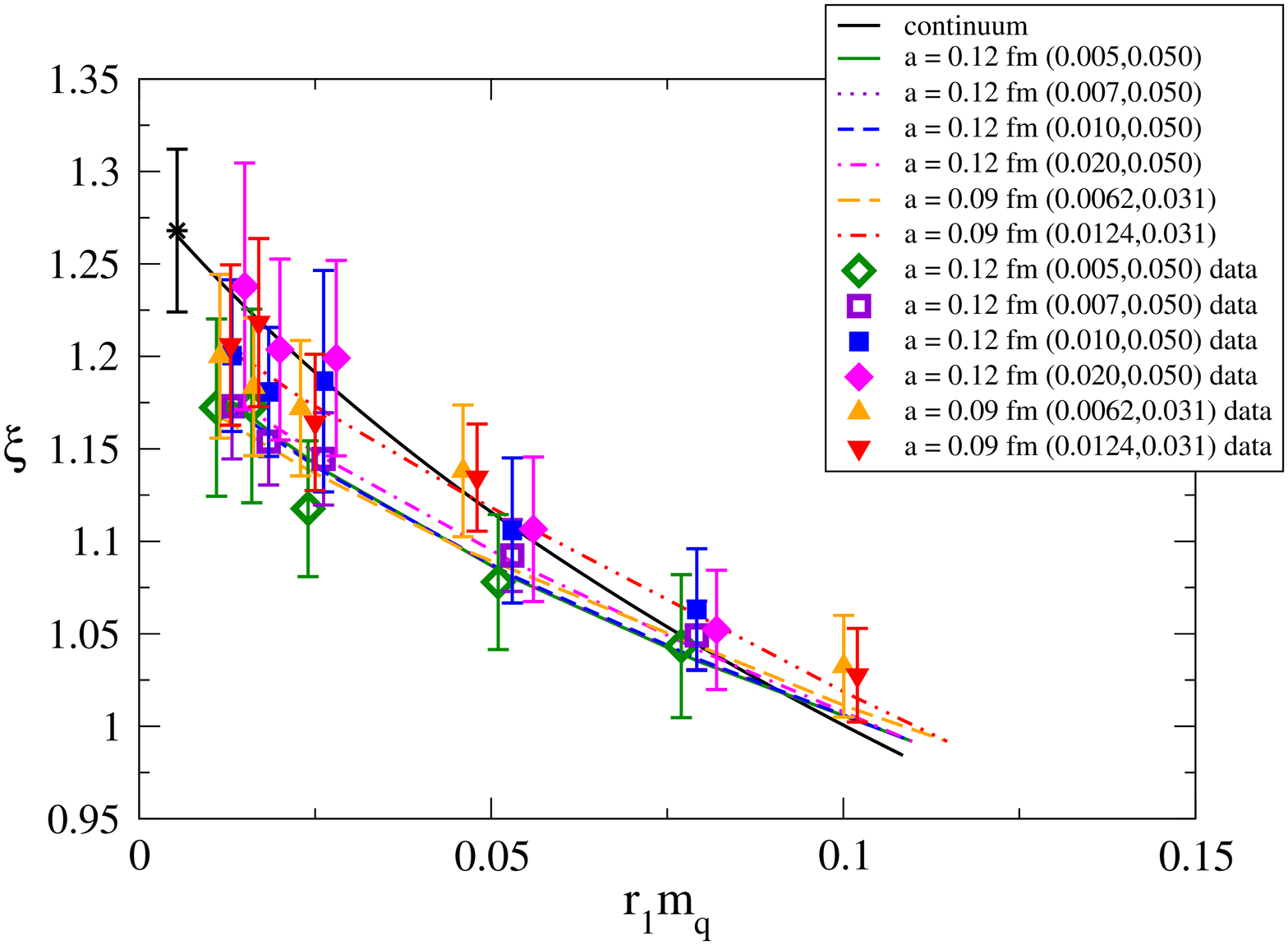}
\caption{\label{fig:xi_NNLO_cf} Fits results using the NNLO rHMS$\chi$PT in
  Eq.~(\ref{eq:xichipt}).  The (black) star is the physical
value of $\xi$ in both plots. The top plot shows only the full QCD data, while
the bottom one shows all the data included in the fits. The (green) squares and
(red) circles and lines in the upper plot represent the $0.12~{\rm fm}$
and $0.09~{\rm fm}$ data and fit results, respectively.
In the bottom plot, each color (symbol) labels a different ensemble.
}

\end{figure}

In Figs.~\ref{fig:xi_NLO_cf} and~\ref{fig:xi_NNLO_cf}, we show the NLO and NNLO 
fit results for~$\xi$ from the direct method 
as a function of the light valence mass in $r_1$ units, $r_1m_q$. The 
top plots in both figures show only the full QCD points, $m_q=m_l$, 
while the bottom plots show  all (partially quenched) 
data included in the fits (see Table~\ref{tab:ensemblesa}). 
The fit curve is the same in both plots of each figure. The black lines show 
the results of the fit in the continuum limit, after the dominant lattice artifacts 
are removed using rHMS$\chi$PT, and after interpolating the physical sea
and valence strange-quark masses to the physical value, as a function of
the valence light-quark mass. The black point is our result for $\xi$ at the physical 
masses, and includes statistical errors.

From the spread of data in the bottom plots of both figures (same data),  
one can see that the light sea-quark mass dependence 
is mild; all different sea-quark masses (squares or triangles at a particular axis 
value) agree within one statistical $\sigma$. The discretization errors are also 
small, as can be seen in both the data and the extrapolation lines in the upper 
plots.

We obtain fits that match the data well and have good $\chi^2/{\rm d.o.f.}$ 
with only the inclusion of NLO terms, as shown in Fig.~\ref{fig:xi_NLO_cf}. 
When we add the NNLO terms, the central values for $\xi$ are also within one 
statistical $\sigma$,  although errors are significantly larger. 
This is to be expected, since the NNLO LECs are poorly known. Related to 
this is the fact that the $\chi^2/{\rm d.o.f.}$ for the NLO fits are 
larger than for the NNLO fits. At NNLO, we are
including extra degrees of freedom with large prior widths that are 
poorly determined by the fit, so, in practice, we are dividing the same $\chi^2$ by a
larger number of degrees of freedom. In fact, the NNLO fits seem to give a slightly 
better description of the data, as can be seen in the full QCD plots. The chiral
extrapolation for $\xi$ is also milder in the NNLO case.
Based on these arguments and, as mentioned above, 
the fact that direct and indirect fits agree better at NNLO, 
we choose the direct NNLO fit for our central value and
statistical error. The systematic error associated with our choice of
fit  function is discussed in Sec.~\ref{chiralerror}. 
  
\section{Error Analysis}

\label{sec:Errors}

In this section, we discuss all  sources of systematic uncertainty 
affecting our calculation of $\xi$. The systematic errors have to be added 
to the statistical uncertainty listed in Table~\ref{tab:directfit_cf}, 
which also encompasses our imperfect 
knowledge about chiral parameters such as $g_{B^*B\pi}$ and $\delta'_{V,A}$.  

\subsection{Heavy-quark mass uncertainty}

The mixing parameters depend on the $b$ quark mass used in our simulations 
through the hopping parameter $\kappa_b$, which is tuned so that the 
kinetic meson mass, $M_2$, agrees with experiment. The dispersion 
relation for a heavy particle  can be written for low-momentum as 

\be
E(\bm{p})=M_1+\frac {\bm{p}^2} {2M_2}-\frac{a^3W_4}{6}\sum_j p_j^4
-\frac{(\bm{p}^2)^2}{8M_4^3}+\order(\bm{p}^6),
\ee
where $\kappa_b$ enters into the definitions of $M_1$ and $M_2$~\cite{Fermilab}.  
$W_4$ and the deviation of $M_4$ from $M_2$ capture lattice artifacts. 
We calculated two-point functions for the pseudoscalar and vector mesons at 
several momenta and extract the energy, $E(\bm{p})$, for each particle at 
each momentum. A fit to the dispersion relation then determines $M_2$  
and the spin average  of the results is taken.  The $\kappa_b$  value
is then adjusted until $M_2$ agrees with the spin-averaged $B_s$ meson mass. 

In this work, we use the values for $\kappa_b$ on the $a\approx0.12~{\rm fm}$  
and $a\approx0.09~{\rm fm}$ ensembles tuned this  way in Ref.~\cite{tuning10}. 
The error in the determination translates into a systematic error in the 
mixing matrix elments. 
However, in the ratio $\xi$ the effect of the uncertainty is minimal, since 
the corrections go in the same direction in both denominator and numerator, 
and, thus, largely cancel. In addition, most of the remaining dependence is 
encoded in the decay constants rather than in the bag parameters, which are 
very insensitive to the exact values of the quark masses. 
In Ref.~\cite{decayconstants2011}, we studied the decay constants 
with the same choice of actions, parameters, and configurations as here. 
We expect systematic errors to be very similar in both analyses. 
We therefore adopt the error due to the uncertainty in the $b$ quark mass 
obtained in Ref.~\cite{decayconstants2011} for the ratio of decay constants 
$f_{B_s}/f_{B_d}$, namely, 0.4\%, 
as a good estimate of this systematic error for~$\xi$.

\subsection{Higher-order effects in the perturbative matching}

\label{sec:errormatching}

The most straightforward and conservative way to estimate the
effects of the missing higher order terms in the perturbative matching is
to assume two-loop coefficients of order 1 and to multiply the central value by
$\alpha^2_s=\alpha^2_V(2/a)$. This estimate gives an error $\sim 5\%$
for $f_B\sqrt{B_B}$ on the $a \approx 0.12~{\rm fm}$ lattices and $\sim 3.6\%$
on the $a \approx 0.09~{\rm fm}$ lattices, becoming the main source of 
uncertainty for this quantity \cite{lattice09}.
If there was no mixing between $\langle {\cal O}_1\rangle$ and $\langle 
{\cal O}_2\rangle$ under renormalization, there would be an exact cancellation of 
the renormalization coefficients for the ratio $\xi=(f_{B_s}\sqrt{B_{B_s}})
/(f_{B_d}\sqrt{B_{B_d}})$, as long as the valence light-quarks are taken to be 
massless in the renormalization calculation. The mixing under renormalization 
prevents this exact cancellation from happening, but the renormalization 
corrections in the ratio are 
still largely suppressed, by a factor of $\langle {\cal O}_2^s\rangle/\langle 
{\cal O}_1^s\rangle-\langle {\cal O}_2^d\rangle/\langle {\cal O}_1^d\rangle$,  
with respect to those for a single matrix element. We estimate this suppression 
factor via the ratio $(m_s-m_{d})/\Lambda_{{\rm QCD}}$ and multiply the 
perturbative error for $f_B\sqrt{B_B}$ given above by it.  
As a result, the perturbative matching uncertainty 
for $\xi$ from this estimate is 0.2--0.5\% (for $\Lambda_{\rm QCD}=700~{\rm MeV}$). 

The ratio $\xi$ changes by $0.2\%$ when the  one-loop renormalization 
is omitted entirely, supporting our power-counting argument. 
Another way of estimating $\order(\alpha_s^2)$ effects is by varying the
scale $q^*$ at which $\alpha_V$ is evaluated.
If we change $q^*$ from our central value of $2/a$  to $1/a$ and $3/a$
we find that  the extrapolated $\xi$ changes between $0.2$--$0.4\%$. 

Since the initial estimate yields the largest uncertainty, $0.5\%$, we take this as 
the error associated with the missing higher order terms in the perturbative
renormalization. Hence, this source of uncertainty is 
subdominant in our determination of $\xi$.

\subsection{Mixing with wrong-spin four-fermion operators}

\label{sec:wrongspin}

As mentioned in Sec.~\ref{sec:chiralfits}, there are contributions at NLO 
in rHMS$\chi$PT originating from the mixing of $\langle\mathcal{O}_1\rangle$ 
with the matrix elements of four-fermion operators of different spin and taste.
We have omitted these contributions from our chiral fits, because we 
discovered these terms after this stage of the analysis was complete. 
From Figs.~\ref{fig:xi_NLO_cf} and~\ref{fig:xi_NNLO_cf}, one can see that the 
effect of the wrong-spin mixing is unlikely to be very large, perhaps being 
mostly absorbed into the LECs.

We cannot include the effects of the wrong-spin contributions, because they 
require the matrix elements of $\mathcal{O}_3$, which we have not computed here.
Fortunately, however, we have started a more comprehensive analysis of 
$B$-$\bar{B}$ mixing on a larger set of higher-statistics ensembles, 
including $\mathcal{O}_3$. We have added the wrong-spin operators to that 
analysis and find that their inclusion tends to increase the slope of the 
continuum extrapolated chiral fit function for $\langle\mathcal{O}_1\rangle$ 
and, hence, $\xi$. For example, taking priors and widths similar to those in 
Table~\ref{tab:ChPTpar}, we find a 2\% increase in $\xi$, while for other 
reasonable choices of the priors the variation is not larger than 3.2\%.
We add a 3.2\% systematic error to account for the missing terms in our 
chiral extrapolation functions.

\subsection{Chiral-extrapolation systematics and light-quark discretization}

\label{chiralerror}

The errors due to the choice of fit ansatz and light-quark discretization effects
cannot be disentangled, because every fit ansatz necessarily treats the discretization 
errors differently. So any estimate of the systematic uncertainty associated 
with the choice of ansatz also accounts for the light-quark discretization errors 
left over after removing the dominant ones using rHMS$\chi$PT.

\begin{figure}
\includegraphics[angle=-90,width=0.60\textwidth]{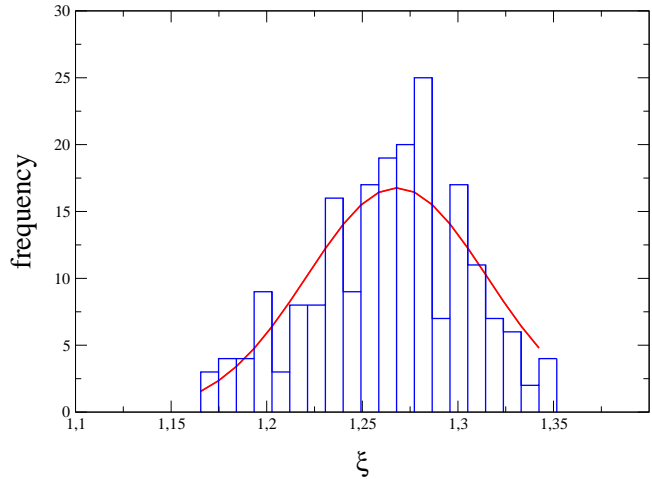}
\caption{Histogram of the distribution of values of $\xi$ obtained from the 200
bootstrap samples. The red line is a Gaussian distribution corresponding 
$\xi=1.268\pm0.047$ (the NNLO result with augmented errors as explained in the text).
\label{fig:bootsxi}}
\end{figure}

In Fig.~\ref{fig:bootsxi}, we show the distribution of values for $\xi$ obtained
with the NNLO direct fit for the 200 bootstrap samples analyzed. 
We check that 200 bootstrap samples is enough to 
obtain a (nearly) Gaussian distribution, as can be seen in the plot. 
With the goal of testing our choice of functional form and the error associated
 with  the truncation of the chiral series, we perform 
fits with only two of the three NNLO terms, omitting each one in turn.
All fits give good values of $\chi^2/{\rm d.o.f.}$ and p value.
The values of $\xi$ obtained are scattered around the distribution in 
Fig.~\ref{fig:bootsxi} but always within $1.5$ statistical $\sigma$ of 
the central value. This consistency, together with 
the fact that the NNLO LECs are not well determined by the fit, indicates 
that the statistical error already accounts for 
the possibility of having one of the unknown constants equal or close  
to zero. If we inflate and symmetrize our statistical errors 
to $\pm0.047$, we cover the spread of results from the different fitting 
functions tried (including the NLO one). We take this value as our estimate of 
both statistical and chiral systematic uncertainties.

An alternative way of estimating the uncertainty in the truncation of the chiral 
series and the fitting function would be taking the difference between the NLO and the 
NNLO fits results. If we add this difference with the statistical errors in 
Table~\ref{tab:directfit_cf} in quadrature, the uncertainty would be slightly smaller 
than the $\pm0.047$ we are taking as our estimate of these two sources of error.

In the rest of this section, we list the errors associated with the 
uncertainty of several input parameters used in the continuum and chiral 
fits, that typically can be estimated by varying the inputs and redoing 
the fits.

\subsubsection{Light-quark mass uncertainty}

The physical values of the light-quark masses used for the extrapolations and 
interpolations for $\xi$ are determined by the MILC collaboration 
\cite{aubin-2004-70,Bernard:2007ps}. They are obtained by making the charged 
pions and kaons take on their physical values after removal of electromagnetic 
effects and are listed in Table~\ref{tab:lmasses}.  

\begin{table} [t]
    \centering
\caption{\label{tab:lmasses} Input for the physical 
light-quark masses used in the chiral extrapolations.  These values were
determined by the MILC collaboration \cite{aubin-2004-70,Bernard:2007ps}.
Physical values are found from chiral fits that have been 
extrapolated to the continuum, but masses are still in units of the 
$a\approx 0.09~{\rm fm}$ lattice spacing.}
\begin{tabular}{ll}
\hline\hline
Quantity & Physical \\
\hline
$am_s\times 10^2$ & 2.72 (8)\\ 
$a\frac{(m_u+m_d)} 2 \times 10^3$& 0.997(35)\\
$am_d\times 10^3$& 1.40(6)\\
\hline\hline
\end{tabular}
\end{table}

The error on $\xi$ due to the light-quark mass uncertainties is obtained 
by individually varying each quark mass within this uncertainty and repeating the 
preferred chiral fit and extrapolation. The central values arrived at using each 
mass variation are compared to the results of the fit which used the central 
values for the masses, and their differences are 
added in quadrature.  This gives a total systematic error due to the 
light-quark mass uncertainties of 0.5\% for~$\xi$.

\subsubsection{Uncertainty in the scale $r_1$}

The value of $r_1$ used in this analysis to convert from lattice to physical 
units, as described in Sec.~\ref{sec:parmCHPT}, is $r_1=0.3117(22)$~fm. 
The results discussed in previous sections are obtained by fixing $r_1$ to its 
central value. In order to estimate the uncertainty due to the error in 
$r_1$ we change $r_1$ by $\pm0.0022$~fm and all parameters that depend
 on the physical $r_1$ are
appropriately adjusted.  The uncertainty in scale gives a systematic
error of 0.2\%, which is very small due to the fact that $\xi$ is a dimensionless
quantity and the scale only enters in the normalization to lattice units 
of the chiral corrections ($1/(f_\pi r_1)^2$) 
and indirectly via the tuning of the quark masses.

\subsection{Heavy-quark discretization effects}

The discretization errors associated with our choice of heavy-quark action to 
simulate the bottom quarks can be described in terms of the difference in the 
lattice and continuum Wilson coefficients of higher dimension operators in the 
HQET expansion. Those come from two sources: the mismatch between continuum 
and lattice in the Lagrangian and the mismatch in the four-fermion operators whose 
matrix elements yield $f_B\sqrt{B_B}$ and, thus, $\xi$. 
For a particular operator $Q_i$ the error can be 
written in terms of the usual power counting magnitudes times 
functions that reflect the particular $m_0a$ 
dependence of the action \cite{Kronfeldlat2003,RuthBtopi}
\ba\label{disc_error}
\texttt{error}_i = z_i f_i(m_0a)\left(a\Lambda_\mathrm{QCD}\right)^{s_i}\, ,
\ea
where $s_i={{\rm dim}\,Q_i-4}$ for Lagrangian operators $Q_i$ 
of dimension 4 and 5, and $s_i={{\rm dim}\, Q_i-6}$ for four-fermion 
operators $Q_i$ of dimension 7 and 8, and the $z_i$ are constants. 
The functions $f_i(m_0a)$ can be deduced from references 
\cite{Fermilab,oktaykronfeld08} and were discussed in detail in 
\cite{RuthBtopi} for the form factors parametrizing $B\to\pi l\nu$ decays 
and in \cite{decayconstants2011} for heavy-light decay constants. A detailed study of 
these corrections for the matrix elements of all the operators contributing 
to neutral $B$ mixing in the SM and beyond will be presented 
elsewhere~\cite{B0perturbative}. 
Here we only summarize the sources of the different 
corrections for $\langle {\cal O}_1\rangle$ and $\xi$. The explicit form of 
the different functions $f_i(m_0a)$ can be found in Appendix~\ref{app:HQDerrors}.  

From the Lagrangian, there are 
$\order(a^2)$ errors and $\order(\alpha_s a)$ errors which are identical to those in 
Eqs.~(A12) and~(A19) in Ref.~\cite{decayconstants2011}. They are 
proportional to the functions $f_E(m_0a)$ in Eq.~(\ref{eq:fE}) and $f_B(m_0a)$ in
 Eq.~(\ref{eq:fB}) of Appendix~\ref{app:HQDerrors}, respectively. 

From the four-fermion operators, we have $\order(a^2)$ errors coming from higher
order corrections to the rotation relation Eq.~(\ref{eq:rotation}). They
are generated by the mismatch between lattice and continuum coefficients
of the operators $\bar q \Gamma \bm{D}^2 b$, $\bar q \Gamma i \bm{\Sigma}
\cdot\bm{B} b$ and $\bar q \Gamma  \bm{\alpha}\cdot\bm{E} b$
in the same way as in Eqs.~(A13) and~(A14) of Ref.~\cite{decayconstants2011}, 
but with an extra overall factor of two due to the fact that we have 
two heavy fields in our 
leading-order operator, not one. These corrections are proportional to the 
functions $f_X(m_0a)$ and $f_Y(m_0a)$ in Eq.~(\ref{eq:fXandfY}) of
 Appendix~\ref{app:HQDerrors}, respectively.

The last contribution, and the least straightforward, comes from 
the $\order(\alpha_s a)$ corrections to the four-fermion operators. 
In principle, to subleading order, there 
are a basis of twelve new local operators in the effective Hamiltonian. 
However, using symmetry constraints, Fierz transformations and rewriting  
some combinations as total derivatives, only five independent operators 
remain~\cite{lattice09,KilianMannel92}. Because we separate the temporal 
and spatial parts of the operators through this analysis, the total 
temporal and spatial components of those five operators should be compared 
with the temporal and spatial parts of the leading operators. 
As explained in Ref.~\cite{B0perturbative}, this produces a difference of 
$3f_3(m_0a)$, where $f_3$ is given in Eq.~(\ref{eq:f3}).

In Table~\ref{tab:HQDE}, we list all the contributions together with the 
functions $f_i$, and the proportionality constant $z_i$ in 
Eq.~(\ref{disc_error}). We also list the numerical values of the 
different contributions to the heavy-quark discretization error 
in percentage for $f_B\sqrt{B_B}$. In order to get the numerical results, 
we use $\Lambda_\mathrm{QCD} = 700~{\rm MeV}$ and 
$\alpha_s=\alpha_V(2/a)$ listed in Table~\ref{tab:rhos}. The final heavy-quark  
discretization error for $f_B\sqrt{B_B}$ is 1.3\% for the $a\approx 0.09~{\rm fm}$ 
ensembles and 2.1\% for the $0.12~{\rm fm}$ ones. 
\begin{table}[t]
\caption{Heavy-quark discretization errors given in percent.\label{tab:HQDE}}
\begin{tabular}{ccccc}
\hline
\hline
Contribution & $f_i$ & $z_i$ & \begin{tabular}{c}Error $f_B\sqrt{B_B}$ (\%)\\
(coarse,fine)\\\end{tabular} & \begin{tabular}{c}Error $\xi$ (\%)\\(coarse,fine)\\
\end{tabular} \\
\hline
 $\order(a^2)$ Lagrangian & $f_E$ & 2 & (0.28,0.16) & \\
$\order(\alpha_s a)$ Lagrangian & $f_B$ & 2 & (0.96,0.58) & \\
$\order(a^2)$ Operator & $f_X$ & 4 & (1.29,0.74) & \\
                       & $f_Y$ & 2 &  (0.23,0.18) & \\ 
$\order(\alpha_s a)$ Operator & $f_3$ & 3  & (1.32,0.75) & \\
\hline
\multicolumn{3}{c}{Total error} & (2.1,1.3) & (0.2,0.1)\\
\hline
\hline
\end{tabular}
\end{table}

These errors largely cancel in $\xi$. The effect of the cancellation on the 
error can be estimated by multiplying the errors in $f_B\sqrt{B_B}$ by 
a factor of $(m_s-m_d)/\Lambda_\mathrm{QCD}$ which gives a final 
heavy-quark discretization error for $\xi$ of $0.2\%$ for the 
coarse lattice and $0.1\%$ for the fine lattice. This agrees well with the 
estimate of this type of error for the ratio 
$f_{B_s}/f_B$~\cite{decayconstants2011}, $\sim 0.3\%$, using a very similar 
set of data and statistics. The strategy  
followed in Ref.~\cite{decayconstants2011} differs from the one described here. 
In that paper, terms of the form in (\ref{disc_error}) were directly added to 
the chiral and continuum extrapolation fitting functions with a coefficient of 
order one to be determined by the fit. Ultimately, we 
would like to employ that strategy also for $B^0$-$\bar B^0$ mixing studies. 
For this work, however, we simply take the larger estimate from the ratio 
$f_{B_s}/f_B$ as our estimate of the uncertainty in $\xi$ due to heavy-quark 
discretization errors.

\subsection{Finite volume corrections}

In order to evaluate the finite volume corrections in our calculation,   
we follow the prescription in Refs.~\cite{aubin-2006-73} and~\cite{bernard-2002-65}. 
The MILC lattices are large enough in the time direction that it 
can be treated as infinite to a very good approximation, so we are   
interested in corrections due to finite spatial volume only. 
They are estimated by replacing infinite-volume integrals in the 
chiral expression with finite sums over the spatial momentum.

Including finite volume corrections in the chiral expressions 
and redoing the fits reveals negligible errors, $<0.1\%$.

\subsection{Tuning of the tadpole parameter \boldmath$u_0$}

The tadpole improvement factor $u_0$ is a parameter of the gauge and asqtad staggered (sea) 
quark action and is determined from the fourth root of the average plaquette. The tadpole 
improvement factor also enters into the valence light and heavy quark actions. On the 
$a \approx 0.09$ fm ensembles, the valence quarks are generated with the same values of 
$u_0$ as the sea. However, on the $a \approx 0.12$ fm ensembles, the valence quark actions 
use values of $u_0$ obtained from the average link in Landau gauge instead. The differences 
between the values of $u_0$ obtained with the two methods is around 3-4\%. 

The effect on $f_{B_s}/f_B$ of the mismatch between $u_0$ values in 
the valence and the sea sectors of the $a \approx 0.12$~fm 
ensembles was estimated to be $<0.1\%$ in 
Ref~\cite{decayconstants2011}. Since this is much smaller than the 
errors due to statistics, chiral fits, and continuum and
chiral extrapolation, we take this estimate as our error on $\xi$.

\section{Discussion of results and future improvements}

\label{sec:Results}

\begin{table}[t]
    \centering
\caption{\label{tab:errorbudget} Complete error budget and total error for
the $B^0$ mixing parameter $\xi$. All errors are given in percent.}
\begin{tabular}{lc}
\hline
 \hline
Source of uncertainty &  Error (\%) \\
\hline
Statistics $\oplus$ light-quark disc. $\oplus$ chiral extrapolation & $3.7$\\
Mixing with wrong-spin operators & $3.2$ \\
Heavy-quark discretization & $0.3$ \\
Scale uncertainty ($r_1$) & $0.2$ \\
Light-quark masses & $0.5$ \\
One-loop matching & $0.5$ \\
Tuning $\kappa_b$ & $0.4$ \\
Finite volume & $0.1$ \\
Mistuned coarse $u_0$ & $0.1$ \\
\hline
Total Error & 5.0  \\
\hline\hline
\end{tabular}
\end{table}
The error budget for the $SU(3)$-breaking mixing parameter $\xi$ described in the 
previous sections is summarized in Table~\ref{tab:errorbudget}. For the first 
error in the table we prefer not to attempt to disentangle the 
statistical, light-quark discretization, and chiral extrapolation errors 
since, as explained in Sec.~\ref{chiralerror}, the lack of 
knowledge about the LECs at NNLO makes a reliable separation  impossible.
Our final result is   
\ba\label{eq:result}
\xi = 1.268\pm 0.063\, .
\ea
The total uncertainty is dominated by the combined statistical, 
light-quark discretization, and chiral extrapolation error, 
and the uncertainty associated with the wrong-spin operators in the 
chiral-continuum extrapolation.

Combining our result in Eq.~(\ref{eq:result}) with the averages of the 
experimentally measured values of the mass differences $\Delta M_d = 
(0.507\pm0.004)~{\rm ps}^{-1}$~\cite{PDG2011} 
and $\Delta M_s =(17.69\pm0.08)~{\rm ps}^{-1}$~\cite{DeltaMsHFAG}, 
and the meson masses $M_{B_s^0}=(5366.0\pm0.9)~{\rm MeV}$ 
and $M_{B_d^0}=(5279.5\pm0.5)~{\rm MeV}$~\cite{PDG2011}, we quote a value 
for the ratio of the CKM matrix elements
\ba 
\left\vert\frac{V_{td}}{V_{ts}}\right\vert = 0.216\pm0.011\,,
\ea
assuming no new physics in $B^0_{(s)}$-$\bar{B}^0_{(s)}$ mixing. 
The error includes the uncertainties in the 
$B$-meson masses and mass differences but it is strongly dominated by the 
error in $\xi$. 

We can also take our result for $\xi$ and combine it with the value of the decay 
constant ratio $f_{B_s}/f_{B_d}=1.229\pm0.026$ calculated by our 
collaboration~\cite{decayconstants2011} to determine the ratio of bag parameters
\ba\label{eq:ratiobag}
\frac{B_{B_s}}{B_{B_d}} = \xi^2\left(\frac{f_{B_d}}{f_{B_s}}\right)^2 = 1.06\pm0.11\,.
\ea
The two results for $\xi$ and $f_{B_s}/f_{B_d}$ are correlated, but the 
statistical analyses were done indepedently so we cannot include the correlations in 
calculating the uncertainty in the ratio of bag parameters. Therefore the error shown 
in the result of Eq.~(\ref{eq:ratiobag}) is overestimated. However, as part of the 
future work, we plan to perform a common analysis of 
matrix elements and decay constants, from which we will be able to account for 
correlations in extracting the value of the bag parameters and thus greatly reduce 
the error in (\ref{eq:ratiobag}). We will do the same for the individual bag 
parameters corresponding to all the operators in the basis in (\ref{susybasis}).

Our result for $\xi$ in Eq.~(\ref{eq:result}) is in good agreement with 
the HPQCD value obtained in Ref.~\cite{HPQCB0mixing}, $\xi=1.258(33)$. 
Note, however, that HPQCD did not estimate the effects of the wrong-spin 
operators that appear in the complete NLO chiral expression, so the full 
error in their result may be somewhat bigger than what was quoted. 
The agreement of these two determinations of $\xi$ provides an excellent check 
of the methodology and systematic error study in both analyses. In addition,
it helps to increase the confidence in the robustness of lattice results for a
parameter of great importance in phenomenological studies. In this article, we 
have established and tested the methodology to apply to broader 
studies of $B^0$ mixing with the same lattice formulations for light and 
bottom quarks as used here.

Statistical errors could be reduced significantly by expanding the analysis to 
include the full set of available configurations (approximately 2000) at each 
of the $a\approx 0.12~{\rm fm}$ and $a\approx 0.09~{\rm fm}$ ensembles. 
The current runs of our collaboration on the
extended ensembles are also implementing sources located at a random spatial and 
time location to reduce further the statistical errors. We expect a reduction 
of the statistical errors by about a factor of two. 

The other dominant error of our calculation, the omission 
in the rHMS$\chi$PT analysis of terms generated by wrong-spin operators, 
will be eliminated when a complete analysis is done with the full rHMS$\chi$PT 
expressions~\cite{B0sCHPT}. A result for $\xi$ that properly includes the 
wrong-spin terms requires the calculation of the continuum matrix elements not 
only of the operator ${\cal O}_1$ 
as we have done in this work, but also of ${\cal O}_2$ and ${\cal O}_3$, and 
simultaneous chiral and continuum extrapolations of all three matrix elements.

The discretization errors, related to both heavy and light quarks, 
will be reduced in a straightforward way by simulations at smaller 
lattice spacing, {\it i.e.}, on the $a\approx 0.06~{\rm fm}$ and 
$a\approx 0.045~{\rm fm}$ 
MILC lattices. The reduction of both statistical and discretization 
errors will also yield cleaner and more accurate continuum 
and chiral extrapolations. Including data at smaller
lattice spacings will also reduce the uncertainty associated with the perturbative 
matching from the reduction of $\alpha_s=\alpha_V(2/a)$. Although not relevant 
for the reduction of the total error in $\xi$, this will be important in the 
determination of the matrix elements $\langle {\cal O}_i\rangle$ themselves.

Similarly, although the uncertainty associated with heavy-quark 
discretization effects is a subdominant source of error in the determination 
of $\xi$, it is one of the main errors in the determination of 
$\langle {\cal O}_i\rangle$~\cite{lattice09}. 
In order to have a more reliable, data-driven estimation of these effects, 
in our on-going  analyses we plan to employ the strategy used in 
Ref.~\cite{decayconstants2011}, in which terms 
like the ones in (\ref{disc_error}) are included in the chiral-continuum 
extrapolation fitting form with free parameters to be determined from the fit. 

Our new analysis, which incorporates the improvements mentioned above,  
includes the study of the matrix elements of all five operators that contribute to 
$H_{eff}^{\Delta B=2}$~\cite{B0proceedings2011}. 
This will allow not only the precise SM determination of  $\Delta M_{s,d}$, 
$\Delta\Gamma_{s,d}$, and $\xi$, but will also provide the nonperturbative 
inputs needed to put constraints on BSM models using experimental data 
on $B^0$ mixing and related observables.

\begin{acknowledgments}
Computations for this work were carried out with resources provided by
the USQCD Collaboration, the Argonne Leadership Computing Facility,
the National Energy Research Scientific Computing Center, and the
Los Alamos National Laboratory, which are funded by the Office of Science of the
U.S. Department of Energy; and with resources provided by the National Institute
for Computational Science, the Pittsburgh Supercomputer Center, the San Diego
Supercomputer Center, and the Texas Advanced Computing Center, which are funded
through the National Science Foundation's Teragrid/XSEDE Program.
This work was supported in part by the U.S. Department of Energy under
Grants No.~DE-FC02-06ER41446 (C.D., L.L., M.B.O.),
No.~DE-FG02-91ER40661 (S.G.),
No.~DE-FG02-91ER40677 (C.M.B, R.T.E., E.D.F., E.G., R.J., A.X.K.),
No.~DE-FG02-91ER40628 (C.B.),
No.~DE-FG02-04ER-41298 (D.T.);
by the National Science Foundation under Grants
No.~PHY-0555243, No.~PHY-0757333, No.~PHY-0703296, No.~PHY10-67881 (C.D., L.L., M.B.O.),
No.~PHY-0757035 (R.S.), and No.~PHY-0704171 (J.E.H.);
by the MICINN, Spain, under grant FPA2010-16696 and \emph{Ram\'on y Cajal}
program (E.G.);
by Junta de Andaluc\'{\i}a, Spain, under grants FQM-101, FQM-330, FQM-03048, and
FQM-6552 (E.G.);
by the URA Visiting Scholars' program (C.M.B., R.T.E., E.G., M.B.O.);
by the Fermilab Fellowship in Theoretical Physics (C.M.B.);
and by the Science and Technology Facilities Council and the Scottish Universities
Physics Alliance (J.L.).
This manuscript has been co-authored by employees of Brookhaven Science
Associates, LLC, under Contract No. DE-AC02-98CH10886 with the
U.S. Department of Energy.
Fermilab is operated by Fermi Research Alliance, LLC, under Contract
No.~DE-AC02-07CH11359 with the United States Department of Energy.
\end{acknowledgments}

\appendix

\section{Staggered Chiral Perturbation Theory for \boldmath$B^0$-$\bar{B}^0$ mixing}

\label{app:SCHPT}

In this appendix we describe the functional form we use in the chiral and
continuum extrapolation of the matrix elements $\langle \overline{B}_q^0|               
{\cal O}_1^q|B_q^0 \rangle$. Further discussion, as well as complete NLO
rHMS$\chi$PT expressions for $\langle \overline{B}_q^0|{\cal O}_i^q|B_q^0 \rangle$
with $i=1,\dots,5$ and the corresponding bag parameters can be found in
\cite{B0sCHPT}.

At NLO in rHMS$\chi$PT and at first order in the heavy-quark expansion we use

\bea \label{eq:ChPTO1}
\langle \overline{B}_q^0|{\cal O}_1^q|B_q^0 \rangle & = &
\alpha \left(1 + \frac{{\cal
W}_{q\overline{b}}+ {\cal W}_{b\overline{q}}}{2} + {\cal T}_q + {\cal
Q}_q\right)\nonumber\\
&& + \,L_v m_q + L_s (2m_l +m_h) + L_a a^2\,.
\eea

\noindent $\alpha$, $L_v$, $L_s$, and $L_a$
are constants to be determined from the fits to lattice data. The quantities in script
for the partially quenched 2+1 ($m_u=m_d\ne m_s$) case are

\bea \label{WFCHPT}
{\cal W}_{q\overline{b}} &=& {\cal W}_{b\overline{q}} =
\frac{ig_{B^*B\pi}^2}{f_\pi^2}\Bigg\{\frac{1}{16}\sum_{\mathscr{S},\rho}
N_\rho\,{\cal H}_{q\mathscr{S},\rho}^{\Delta^*\!+\delta_{\mathscr{S}q}}
+\frac{1}{3}\bigg[R^{[2,2]}_{X_I}\big(\{M^{(5)}_{X_I}\}
;\{\mu_I\}\big)\; \frac{\partial{{\cal
H}^{\Delta^*}_{X,I}}}{\partial m^2_{X_I}}  \nonumber
\\ && -\hspace{-3mm}\sum_{j \in                       
\{M^{(5)}_I\}}D^{[2,2]}_{j,X_I}\big(\{M^{(5)}_{X_I}\};\{\mu_I\}\big){\cal
H}^{\Delta^*}_{j,I} \bigg]
+a^2\delta'_{V}\bigg[R^{[3,2]}_{X_V}\big(\{M^{(7)}_{X_V}\}
;\{\mu_V\}\big)\; \frac{\partial{{\cal
H}^{\Delta^*}_{X,V}}}{\partial m^2_{X_V}}   \nonumber
\\ &&
-\hspace{-3mm}\sum_{j \in                                             
\{M^{(7)}_V\}}D^{[3,2]}_{j,X_V}\big(\{M^{(7)}_{X_V}\};\{\mu_V\}\big){{\cal
H}^{\Delta^*}_{j,V}
 \bigg] +\big(V\rightarrow A\big)\Bigg\}}.
\ea

 \bea {\cal T}_{q} &=&
\frac{-i}{f_\pi^2}\Bigg\{\frac{1}{16}\sum_{\mathscr{S},\rho}
N_\rho\,{\cal I}_{q\mathscr{S},\rho}
+\frac{1}{16}\sum_{\rho} N_\rho {\cal
I}_{X,\rho}
+\frac{2}{3}\bigg[R^{[2,2]}_{X_I}\big(\{M^{(5)}_{X_I}\}
;\{\mu_I\}\big)\left(\frac{\partial{\cal I}_{X_I}}{\partial
m^2_{X_I}}\right) \nonumber \\ && -\sum_{j \in 
\{M^{(5)}_I\}}D^{[2,2]}_{j,X_I}\big(\{M^{(5)}_{X_I}\};\{\mu_I\}\big){\cal
I}_{j} \bigg]
+a^2\delta'_{V}\bigg[R^{[3,2]}_{X_V}\big(\{M^{(7)}_{X_V}\}
;\{\mu_V\}\big)\left(\frac{\partial{\cal I}_{X_V}}{\partial
m^2_{X_V}}\right)  \nonumber
\\ &&
-\sum_{j \in                                                                          
\{M^{(7)}_V\}}D^{[3,2]}_{j,X_V}\big(\{M^{(7)}_{X_V}\};\{\mu_V\}\big){\cal
I}_{j}
 \bigg] +\big(V\rightarrow A\big)\Bigg\}, \eea
\bea \label{eq:appAChPTO1}
{\cal Q}_{q} &=&
\frac{-ig_{B^*B\pi}^2}{f_\pi^2}\Bigg\{\frac{1}{16}\sum_{ \rho}\,
N_\rho {\cal H}^{\Delta^*}_{X,\rho}
+\frac{1}{3}\bigg[R^{[2,2]}_{X_I}\big(\{M^{(5)}_{X_I}\}
;\{\mu_I\}\big)\left(\frac{\partial{{\cal
H}^\Delta_{X_I}}}{\partial m^2_{X_I}}\right) \nonumber
\\ && -\sum_{j \in                                                                    
\{M^{(5)}_I\}}D^{[2,2]}_{j,X_I}\big(\{M^{(5)}_{X_I}\};\{\mu_I\}\big){\cal
H}^\Delta_{j} \bigg] \Bigg\}, \eea
In the equations above, the index $\rho$ runs over the taste representation
$(P,A,T,V,I)$ with degeneracies $N_\rho$ ($N_\rho=$1,4,6,4,1, respectively),
and $\mathscr{S}$ runs over the sea flavors $u$, $d$, $s$. The meson $X$ is made of
two light valence quarks $q$, and $m_X$ is its mass.
The functions ${\cal H}$ and ${\cal I}$ are the integrals
defined in Appendix A of~\cite{detmold-2006}. The subscripts on those functions
label the flavor and taste of the meson masses at which they are
evaluated.

The superscript in ${\cal H}$ is the second argument for that function as
defined in~\cite{detmold-2006}. In addition to the hyperfine splitting
$\Delta^*=M_{B^*}-M_B$, it includes a light flavor splitting $\delta_{\mathscr{S}q}$
whenever the light flavor of the vector meson in the loop is different from
the external flavor.

The splitting is  $\delta_{\mathscr{S}q} \equiv M_{B^0_\mathscr{S}} - M_{B^0_q}=        
2\lambda_1\mu (m_\mathscr{S}-m_q)$, where $\lambda_1$ and $\mu$ are
 low energy constants. The constant $\lambda_1$ comes from heavy quark effective
theory, and $\mu$ is defined in (\ref{eq:Mij}).

The residues functions $R_j^{[n,k]}$ and $D_{j,l}^{[n,k]}$ in the
expressions above are defined by \cite{Aubin:2003uc}
\begin{eqnarray}
R^{[n,k]}_j(\{m\},\{\mu\})  &\equiv & \frac{\prod_{a=1}^k
(\mu^2_a-m^2_j)}{\prod_{i\neq j} (m^2_i-m^2_j)}, \nonumber \\
D^{[n,k]}_{j, l}(\{m\},\{\mu\})  &\equiv &
-\frac{d}{dm^2_l}R^{[n,k]}_j(\{m\},\{\mu\}).
\end{eqnarray}
The mass combinations appearing as arguments of these functions in the
2+1 partially quenched theory are
\begin{eqnarray} \{M_X^{(5)}\} &\equiv& \{m_\eta, m_X \}, \nonumber \\
    \{M_X^{(7)}\}&\equiv& \{m_\eta, m_{\eta'}, m_X \}, \nonumber \\
    \{\mu\}&\equiv&\{m_L,m_H\}\, , \end{eqnarray}
where $m_L$ is the meson mass made from $l\bar l$ sea quarks, and $m_H$ is the
meson mass made from $h\bar h$ sea quarks. The tastes of these mesons are 
indicated explicitly in the equations above.

Since we are not including the effects of the hyperfine splitting $\Delta^*$
or the light flavor splittings $\delta_k$ in this work, the functions ${\cal H}$
and ${\cal I}$ appearing in the wave function,
tadpole, and sunset contributions simplify to
\ba
i{\cal H}_{k,\Xi}^0 = -3i{\cal I}_{k,\Xi} = -\frac{3}{16\pi^2}
M_{k,\Xi}^2\,\ln\left(\frac{M_{k,\Xi}^2}{\Lambda_\chi^2}\right)\,.
\ea

\section{Functions parametrizing heavy-quark discretization errors}

\label{app:HQDerrors}

In this Appendix we collect the functions $f_i$ needed in Eq.~(\ref{disc_error})
to estimate the heavy-quark discretization errors affecting our calculation.
For details on the origin of these functions and the effects of higher-dimension
operators in the lagrangian, see~\cite{oktaykronfeld08}.
For further details on the application
to the estimation of heavy-quark discretization errors in $B^0-\bar B^0$ mixing,
see~\cite{B0perturbative}.

\begin{itemize}
\item $\order(a^2)$ errors from the Lagrangian.
\bea \label{eq:fE}
f_E(m_0a)  & = &  \frac{1}{2}\left\lbrack\frac{(1+m_0a)-1}{m_0a(2+m_0a)(1+m_0a)}
-\frac{1}{4(1+m_0a)^2}\right\rbrack
\eea
\item $\order(\alpha_s a^2)$ errors from the Lagrangian
\bea \label{eq:fB}
f_B(m_0a) & = & \frac{\alpha_s}{2(1+m_0a)}
\eea

\item $\order(a^2)$ errors from the four-fermion operator
\bea
f_X(m_0a) & = & \frac{1}{2}\left\lbrack\frac{1}{2(1+m_0a)}
-\left(\frac{m_0a}{2(2+m_0a)(1+m_0a)}
\right)^2\right\rbrack \nonumber\\
f_Y(m_0a) & = &\frac{2+4m_0a+(m_0a)^2}{4(1+m_0a)^2(2+m_0a)^2}\label{eq:fXandfY}
\eea
\item $\order(\alpha_s a^2)$ errors from the four-fermion operator
\bea \label{eq:f3}
f_3(m_0a) =  \frac{\alpha_s}{2(2+m_0a)}
\eea

\end{itemize}

\section{Prior central values and widths for the correlator fits}

\label{app:priors}

In the table below we collect the prior central values and widths used in
the correlator fits described in Sec.~\ref{sec:fitting}. The amplitude parameters 
are defined in Eqs.~(\ref{eqn:CQ}) and~(\ref{eqn:CPS}), and the energy 
differences are defined as $\Delta E_{i+1,i}\equiv a(E_{i+1}-E_1)$. 

  \begin{table} [h]
    \centering
  \caption{\label{tab:priors} The priors with index 0 refer to the ground
state. Superscripts $d$ and $1S$ refer to the local and 1S smeared sources
respectively. Higher energy state priors have indices $i$ and $j$. The prime
in $E_i'$ refers to an opposite parity (oscillating) state. }
  \begin{tabular}{ccc}
  \hline
  \hline
   &  Prior central value & Prior width \\
  \hline
  $ Z^{1S}_0$  & 2.2  & 0.5 \\
  $Z^{1S}_i$  & 0.01 & 0.5 \\
  $Z^d_0$  & 0.45 &  0.45 \\
  $Z^d_i$  & 0.01 & 1 \\
  $O_{00}$  & 0.01 & 0.02 \\
  $O_{ij}$ & 0.01 & 0.1 \\
\hline
  $E_{0}$ ($0.12~{\rm fm}$)  &  1.95 & 0.15 \\
  $E'_0$ ($0.12~{\rm fm}$) & 2.25 & 0.15 \\
\hline
  $E_{0}$ ($0.09~{\rm fm}$) &  1.65 & 0.15 \\
  $E'_0$ ($0.09~{\rm fm}$)  & 1.85 & 0.15 \\
\hline
  $\log\Delta E_{i+1,i}$  & -1.5 & 0.5 \\
  $\log\Delta E'_{i+1,i}$  & -1.5 & 0.5 \\
\hline\hline
  \end{tabular}
\end{table}


\end{document}